\documentclass{article}

\PassOptionsToPackage{numbers, compress}{natbib}
\usepackage[final]{neurips_2023}

\usepackage[utf8]{inputenc} % allow utf-8 input
\usepackage[T1]{fontenc}    % use 8-bit T1 fonts
\usepackage{hyperref}       % hyperlinks
\usepackage{url}            % simple URL typesetting
\usepackage{booktabs}       % professional-quality tables
\usepackage{amsfonts}       % blackboard math symbols
\usepackage{nicefrac}       % compact symbols for 1/2, etc.
\usepackage{microtype}      % microtypography
\usepackage{xcolor}         % colors

\usepackage{algorithm}
\usepackage{algpseudocode}
\usepackage{amsfonts}
\usepackage{amsmath}
\usepackage{amssymb}
\usepackage{bm}
\usepackage{color}
\usepackage{enumitem}
\usepackage{float}
\usepackage{graphicx}
\usepackage{hyperref}
\usepackage{mathtools,soul}
\usepackage{multicol}
\usepackage{multirow}
\usepackage{soul}
\usepackage{subcaption}
\usepackage{xspace}
\usepackage{pifont}
\usepackage{siunitx}

\DeclareMathOperator*{\argmax}{argmax}

\DeclareMathOperator*{\concat}{\parallel}
\newcommand{\cmark}{\ding{51}}%
\newcommand{\xmark}{\ding{55}}%
\newcolumntype{d}[1]{S[table-format=#1]}
\newcommand{\specialcell}[2][c]{%
  \begin{tabular}[#1]{@{}c@{}}#2\end{tabular}}

\title{Enhancing User Intent Capture in Session-Based Recommendation with Attribute Patterns}

\author{%
  Xin Liu$^{1}$\thanks{Work was done during Xin's internship at Amazon. Corresponding author: Xin Liu and Zheng Li. } \quad Zheng Li$^{2}$ \quad Yifan Gao$^{2}$ \quad Jingfeng Yang$^{2}$ \\
  \textbf{Tianyu Cao}$^{2}$ \quad \textbf{Zhengyang Wang}$^{2}$ \quad
  \textbf{Bing Yin}$^{2}$ \quad \textbf{Yangqiu Song}$^{1}$\thanks{Prof. Yangqiu Song is a visiting academic scholar at Amazon.} \\
$^1$Department of Computer Science and Engineering, HKUST \quad $^2$Amazon.com Inc \\
\texttt{\{xliucr,yqsong\}@cse.ust.hk}\\
\texttt{\{amzzhe,yifangao,jingfe,caoty,zhengywa,alexbyin\}@amazon.com}
}

\begin{document}

\maketitle

\begin{abstract}
  The goal of session-based recommendation in E-commerce is to predict the next item that an anonymous user will purchase based on the browsing and purchase history. However, constructing global or local transition graphs to supplement session data can lead to noisy correlations and user intent vanishing. In this work, we propose the Frequent Attribute Pattern Augmented Transformer (FAPAT) that characterizes user intents by building attribute transition graphs and matching attribute patterns. Specifically, the frequent and compact attribute patterns are served as memory to augment session representations, followed by a gate and a transformer block to fuse the whole session information. Through extensive experiments on two public benchmarks and 100 million industrial data in three domains, we demonstrate that FAPAT consistently outperforms state-of-the-art methods by an average of 4.5\% across various evaluation metrics (Hits, NDCG, MRR). Besides evaluating the next-item prediction, we estimate the models' capabilities to capture user intents via predicting items' attributes and period-item recommendations.
\end{abstract}

\section{Introduction}
\label{sec:intro}

\begin{figure}[t]
    \begin{minipage}{0.28\textwidth}
    \centering
    \includegraphics[width=0.90\linewidth]{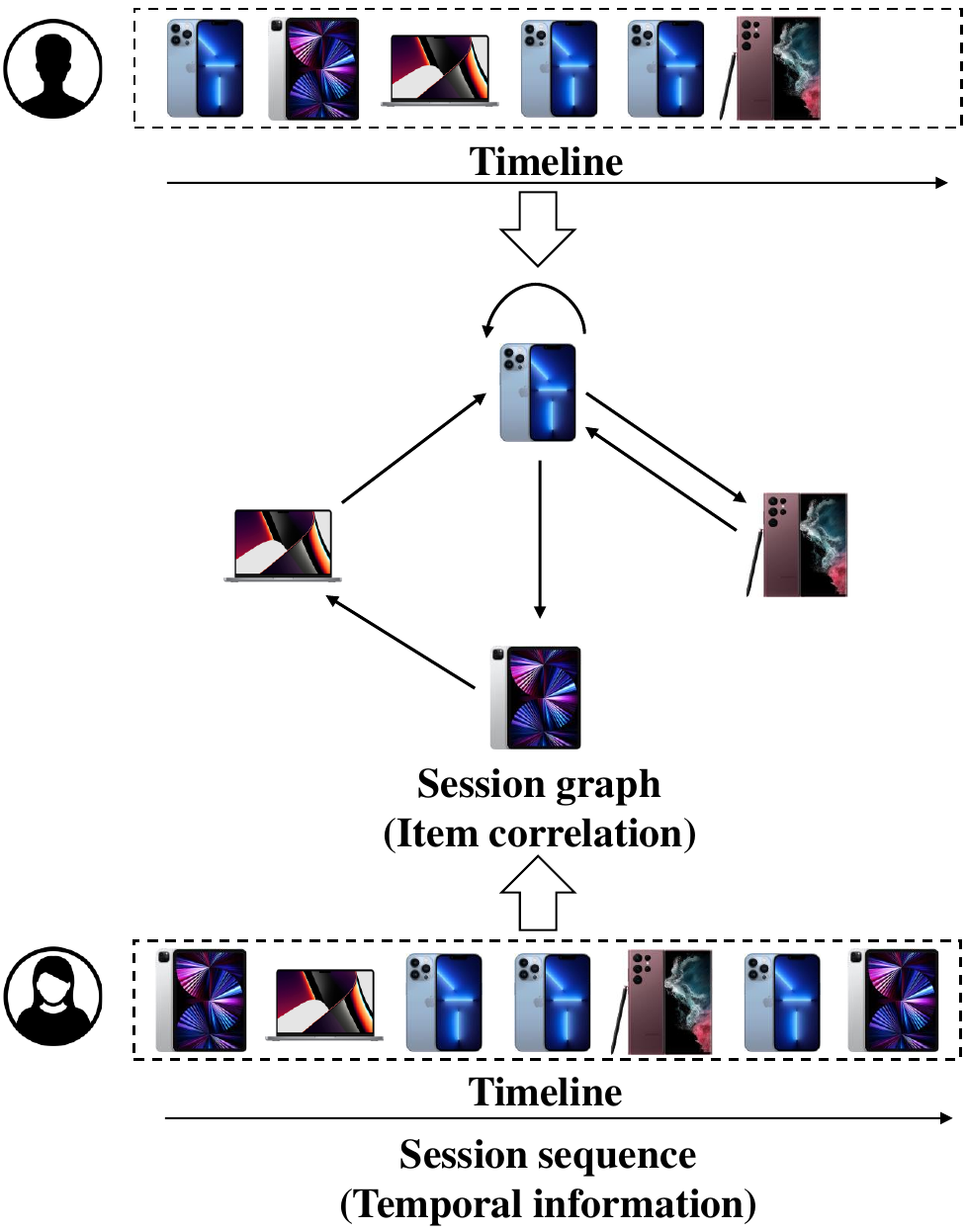}
    \captionof{figure}{Local session graph construction.}
    \label{fig:same_session_graph}
    \end{minipage}
    \begin{minipage}{0.72\textwidth}
    \includegraphics[width=0.95\linewidth]{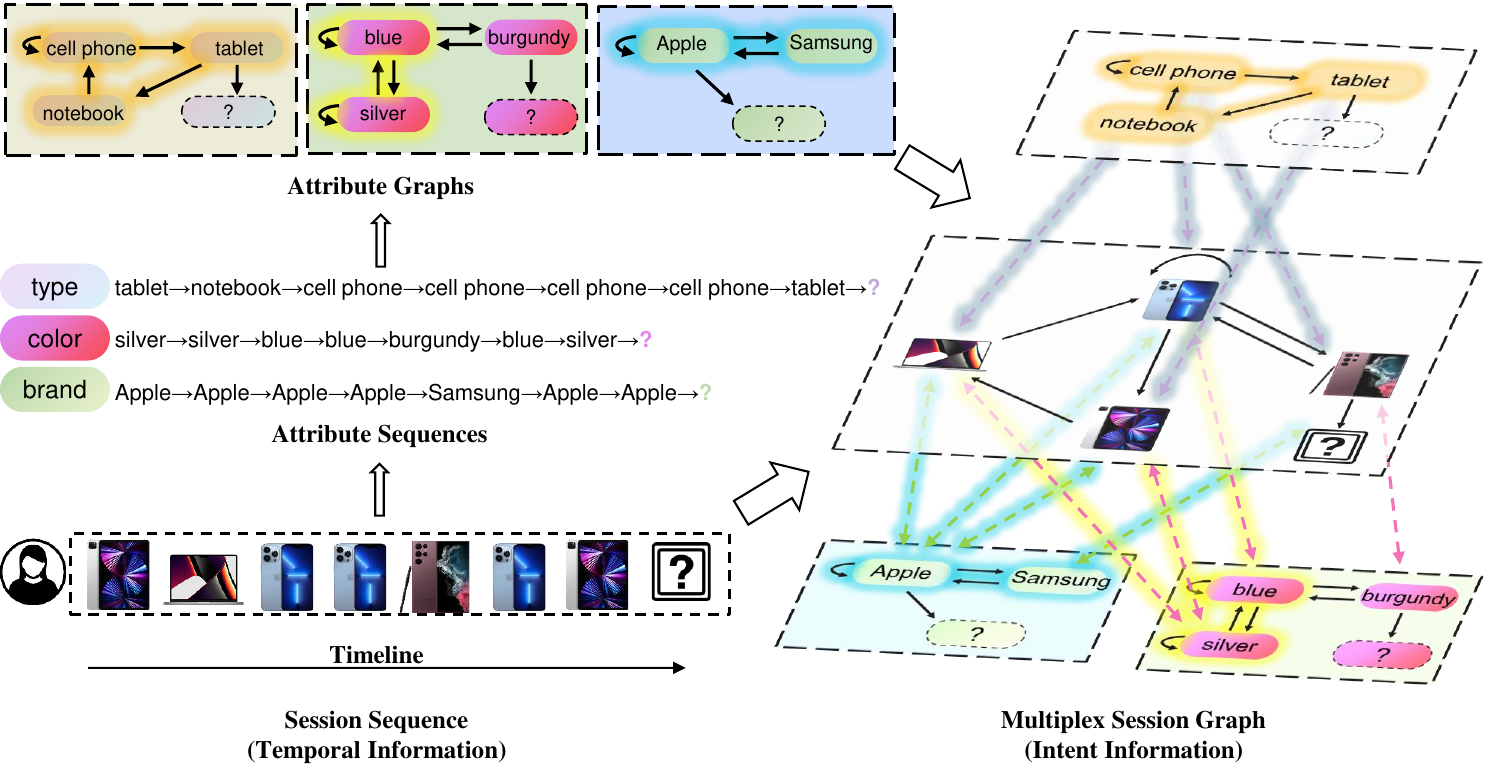}
    \captionof{figure}{
    A session graph enriched by multiplex attribute graphs.}
    \label{fig:multiple_session_graph}
    \end{minipage}
    \vspace{-0.2in}
\end{figure}

With the explosive demand for E-commerce services~\cite{zhang21queaco,jiang2022short,chen22query,huang2023ccgen, yu-etal-2023-folkscope}, numerous user behaviors are emerging. Understanding these historical action records is critical in comprehending users' interests and intent evolution, particularly in a cold-start regime that lacks sufficient context. This has spurred research on session-based recommendations (SBR)~\cite{HidasiKBT15, WuT0WXT19, XiaYYWC021, jin2023amazon} that capture user-side dynamics from a short-time period (namely a session) using temporally historical information.
Numerous SBR algorithms have been proposed, ranging from sequence-based methods~\cite{RendleFS10, HidasiKBT15, LiRCRLM17, LiuZMZ18, WangRMCMR19, ShalabyOAKC22,HouHZZ22} to graph-based methods~\cite{WuT0WXT19, XuZLSXZFZ19, WangWCLMQ20, XiaYYWC021, LinTHZ22,wan2023spatio} for learning dynamic user characterization. However, both lines of methods have their limitations. Specifically, sequence-based methods treat each user behaviors in a session as an action sequence and model the local dependencies inside. This can only capture users' preference evolution via chronological order while failing to identify the complex non-adjacent item correlation, especially when the session length is insufficient to support temporal prediction~\cite{WangZ0ZW022}. To address this issue, graph-based methods adopt a higher perspective by introducing a global item transition graph, which aggregates local session graphs constructed from historical session sequences. Thus, a newly-emerging short session sequence can benefit from the global topology (e.g., global co-occurrences) and representations (e.g., item semantics)~\cite{WangWCLMQ20}. Unfortunately, existing session graph construction ignores temporal signals. Figure~\ref{fig:same_session_graph} shows that two different sessions result in the same session graph, leading to vanishing of user intent variation during sequence-to-graph conversion. 
And such global graphs are fragile due to noise from random clicks.

Besides the global item transition graph, there are other possible solutions for item correlations. One feasible solution involves item-side knowledge. Items can be connected through shared attributes, such as being manufactured by the same company~\cite{ShalabyOAKC22}. We argue that the current use of item-side metadata provides little assistance in SBR models as user intent may change over time. However, such meta-data are still useful from the view of graphs. As illustrated in Figure~\ref{fig:multiple_session_graph}, session attributes can be organized into attribute graphs and anchored to the local session graph to create multiplexes. Such multiplexes provide several intent clues, in addition to item-side correlations. For example, the color pattern {\textit{silver} $\leftrightarrow$ \textit{silver} $\leftrightarrow$ \textit{blue} $\leftrightarrow$ \textit{blue}} reveals the user's color intent, while the brand pattern {\textit{Apple} $\leftrightarrow$ \textit{Apple} $\leftrightarrow$ \textit{Samsung}} implies a potential change in intent. Different sessions can benefit from shared attribute topologies and representations, which can ultimately entail implicit high-order correlations. However, the session graph is essentially a general conditional random field, and optimization becomes intractable due to the large candidate size. Graph neural networks (GNNs) also face challenges due to the possibility of over-smoothing and data noise~\cite{ZhaoA20,LiuCZGN21,abs-2206-12781,wang22nips}.

To alleviate the aforementioned issues, we propose a novel framework called Frequent Attribute Pattern Augmented Transformer (\textbf{FAPAT}) that considers highly frequent attribute patterns as supplementary instead of directly learning multiplex graphs. In traditional graph learning, graphlets (such as triangles, triangular pyramids, etc.) have been proven to be useful features in graph classification and representation learning~\cite{ShervashidzeVPMB09,KriegeJM20}. Therefore, we employ frequent graph pattern mining algorithms to find consequential graphlets and view them as compact hyper-edges (e.g., {\textit{cell phone}, \textit{tablet}, \textit{notebook}} in Figure~\ref{fig:multiple_session_graph}). We then use these attribute patterns as accessible memory to augment session sequence encoding. Before encoding patterns, we use Jaccard similarities to rank and retrieve the most highly correlated patterns, which significantly reduces the graph density and computational cost.
It has been shown that GNNs can estimate the isomorphism and frequency of substructures~\cite{XuHLJ19,LiuPHSJS20,LiuS22}.
Thus, we leverage multi-head graph attention to learn pattern and local session graph representations in the aligned space~\cite{HuDWS20,LiuCSJ22}.
To incorporate temporal signals and capture user intents, we distribute graph representations back to session sequences and use external pattern memory to augment sequence representations via memory attention with relative position bias. Finally, the sequence is fully fused by a transformer block. In other words, graph information is used to aggregate attribute patterns, while temporal actions are used to encode items.

To validate the effectiveness of FAPAT, we conduct extensive experiments on two public benchmark datasets and three real-world large-scale industrial datasets with around 100 million clicks, and experimental results demonstrate significant improvement with an average boost of 4.5\% across various evaluation metrics (Hits, NDCG, MRR).
Compared with baselines, the attribute pattern density can significantly relieve over-smoothing.
Besides, we also extend evaluation to attribute estimations and sequential recommendations to measure the model capability to capture user intents.
Code and data are availiable at \url{https://github.com/HKUST-KnowComp/FAPAT}.

\section{Related Work}
\label{sec:related}
\begin{table}[!t]
    \centering
    \scriptsize
    \setlength\tabcolsep{2pt}
    \caption{Comparison with existing popular methods.}
    \label{tab:comparison}
    % \vspace{-0.1in}
    \begin{tabular}{c|c|c|c|c|c}
        \toprule
         Methods & 
         {Temporal Information} & {History Attention} & {Local Session Topology} & {Global Item Correlation} & {Attribute Association}\\
         \midrule
         FPMC & \cmark & \xmark & \xmark & \cmark & \xmark \\
         GRU4Rec & \cmark & \xmark & \xmark & \xmark & \xmark \\
         NARM & \cmark & \cmark & \xmark & \xmark & \xmark \\
         STAMP & \cmark & \cmark & \xmark & \xmark & \xmark \\
         CSRM & \cmark & \cmark & \xmark & \cmark & \xmark \\
         S3-Rec & \cmark & \cmark & \xmark & \cmark & \cmark \\
         M2TRec & \cmark & \cmark & \xmark & \cmark & \cmark \\
         \hline
         SR-GNN & \xmark & \xmark & \cmark & \xmark & \xmark \\
         GC-SAN & \xmark & \cmark & \cmark & \xmark & \xmark \\
         S2-DHCN & \xmark & \xmark & \cmark & \cmark & \xmark \\
         GCE-GNN & \xmark & \xmark & \cmark & \cmark & \xmark \\
         LESSR & \cmark & \xmark & \cmark & \xmark & \xmark \\
         MSGIFSR & \cmark & \cmark & \cmark & \xmark & \xmark \\
         \hline
         FAPAT & \cmark & \cmark & \cmark & \cmark & \cmark \\
         \bottomrule
    \end{tabular}
    \vspace{-0.2in}
\end{table}

{\flushleft \textit{\textbf{Neural Methods for Session-based Recommendation Systems}}.}
\label{sec:nn_sbr}
Table~\ref{tab:comparison} presents a summary of the distinctions between current neural techniques and our novel FAPAT. The concept of \textit{Temporal Information} implies Markov decision processes with previous histories. The technique of \textit{History Attention} employs attention for learning long-distance sequences. The approach of \textit{Local Session Topology} involves modeling session sequences from the view of session graphs. Lastly, \textit{Global Item Correlation} and \textit{Attribute Association} place emphasis on capturing item-side and attribute knowledge.

{\flushleft \textit{Sequence-based Models}.}
\label{sec:seq_sbr}
FPMC~\cite{RendleFS10} uses first-order Markov chain and matrix factorization to identify sequential patterns of long-term dependencies. However, the Markov-based method usually has difficulty exploring complicated temporal patterns beyond first-order relationships. Recently, neural networks have shown power in exploiting sequential data in SBR tasks, such as GRU4Rec~\cite{HidasiKBT15}. NARM~\cite{LiRCRLM17} extends GRUs with attention to emphasize the user's primary purchase purpose. Similarly, STAMP~\cite{LiuZMZ18} uses an attention-based memory network to capture the user's current interest. These attention-based models separately deal with the user's last behaviors and the whole session history to detect the general and latest interests. But they mainly focus on the user's preference from a temporal view but ignore the item correlations. Pre-training techniques~\cite{ZhouWZZWZWW20} and multi-task learning~\cite{ShalabyOAKC22} also demonstrate effectiveness in injecting item metadata to embeddings and predicting item attributes.
Besides, some recent sequence-based approaches leverage generative pretrained language models to provide explicit explanations for recommendation systems~\cite{cui2022m6,Geng0FGZ22}.

{\flushleft \textit{Graph-based Models}.}
\label{sec:gnn_sbr}
Graph neural networks (GNNs) have recently been explored in SBRs due to the substantial implications behind natural transition topologies. SR-GNN~\cite{WuT0WXT19} adopts a gate GNN to obtain item embeddings over the local session graph and predict the next item with weighted sum pooling, showing impressive results on benchmark data. Some advanced variants have further boosted performance, such as GC-SAN~\cite{XuZLSXZFZ19} with self-attention mechanism and FGNN~\cite{QiuLHY19} with weighted attention graph layers. To acquire further collaborative information, S2-DHCN~\cite{WangRMCMR19} constructs line graphs to capture correlations among neighbor sessions, and GCE-GNN~\cite{WangWCLMQ20} directly applies a graph convolution over the global transitions to aggregate more relevant items for local sessions. However, GNN-based methods still face challenges in capturing temporal signals, filtering noise, and leveraging implicit high-order collaborative information. Other methods, such as LESSR~\cite{ChenW20} and MSGIFSR~\cite{Guo0SZWBZ22}, have also shown significant improvement by building multigraphs and shortcut graphs for session representation learning and user intents from different granularities, respectively.

{\flushleft \textit{\textbf{Pattern Mining for Recommendation Systems}}.}
Pattern mining is an important data mining technique with board applications.
In recommendation systems, sequential pattern mining assists in analyzing customer purchase behaviors through frequent sequential patterns.
Such mining focuses on item patterns with frequencies above a threshold in all sessions, which reduces the diversity of the recommended items~\cite{ChenKWT09}.
Personalized sequential pattern mining~\cite{YapLY12} effectively learns user-specific sequence importance knowledge and improves the accuracy of recommendations for target users.
It can be challenging to generalize to SBR systems when neural networks have already implicitly captured such behavior patterns.
But attribute graph patterns still need to be explored in SBR.
\section{Background and Motivations}

\subsection{Problem Definition}

Session-Based Recommendation (SBR) assumes that users' historical behaviors outside the current session are inaccessible, in contrast to general recommendations. For example, users do not log in due to user privacy and security reasons. SBR predicts the next item that an anonymous user is most likely to click on or purchase based on historical behaviors within a short period. Despite the lack of personal profiles, this universal setting can better reflect the quality of item-side recommendations. Suppose that there are $N$ unique items in the database, and each session is represented as a repeatable sequence of items $S = [v_{1}, v_{2}, \ldots, v_{L}]$, $v_{i} \in \mathcal{V} \ (1 \leq i \leq L)$ represents the $i$-th behavioral item of the anonymous user within session $S$, where $\mathcal{V}$ is the item set collected from overall sessions, and $L$ is the length of the session. Given a session $S$, the goal is to recommend the top-$K$ items $(1 \leq K \leq N)$ that have the highest probabilities of being clicked by the anonymous user.

\subsection{Session Graphs and Transition Graphs}

Session sequence modeling is not always sufficient for SBR as it only reflects transitions from the user side. To account for item correlations, SR-GNN~\cite{WuT0WXT19} converts session sequences to session graphs. Each session graph $\mathcal{G}_{S} = (\mathcal{V}_{S}, \mathcal{E}_{S})$ is a directed graph with node set $\mathcal{V}_{S} \subseteq \mathcal{V}$, consisting of unique items in the session, and edge set $\mathcal{E}_{S}$, recording adjacent relations between two items in session $S$. Edge weights can be normalized by indegrees or outdegrees to model transition probabilities. GCE-GNN~\cite{WangWCLMQ20} extends this graph modeling by merging all session graphs as a global transition graph, which aggregates more item correlations but faces over-smoothing and data noise~\cite{ZhaoA20,LiuCZGN21}.

Instead of modeling global item transitions, we enrich session graphs with attributes and patterns. Assume there are $M$ different kinds of attributes, and the $m$-th attribute type $\mathcal{A}^{(m)} \ (1 \leq m \leq M)$ consists of $|\mathcal{A}^{(m)}|$ possible values. Thus, each item $v \in \mathcal{V}_{S}$ has attribute list $[a^{(1)}_{v}, a^{(2)}_{v}, \cdots, a^{(M)}_{v}]$, where $a^{(m)}_{v} \in \mathcal{A}^{(m)}$ denotes the $m$-th attribute value of $v$. Each session sequence $S$ corresponds to $M$ attribute histories, with the $m$-th attribute sequence denoted as $S^{(m)} = [a^{(m)}_{v_1}, a^{(m)}_{v_2}, \cdots, a^{(m)}_{v_L}]$. Using the sequence-to-graph transform, we convert $S^{(m)}$ to $\mathcal{G}_{S}^{(m)}$, whi is usually denser than the session graph $\mathcal{G}_{S}$. Finally, the $M$ attribute sequences are separately transformed into $M$ attribute session graphs in different property-specific channels, anchored in items $\mathcal{V}_{S}$. Finally, we add edges for item $v_i$ and attribute $a^{(m)}_{v_i}$
to construct a multiplex session graph, preserving attribute values and transitions.
We represent the session graph with attributes as the multiplex $\mathcal{G}_{S}^{\mathcal{A}}$ (as Figure~\ref{fig:multiple_session_graph}).

\subsection{Frequent Pattern Mining}

Frequent pattern mining aims to extract inductive clues from data to comprehend data distributions, which includes two main categories: sequential pattern mining and graph pattern mining. The former is concerned with sequence databases composed of ordered elements, while the latter statistics the important graph structures.
For two sequences $S' = [{v'}_1, {v'}_2, \cdots, {v'}_{L'}]$ and $S = [v_1, v_2, \cdots, v_{L}]$, we refer to $S'$ as the pattern of $S$ if $S'$ is a subsequence of $S$.
Similarly, a graph $\mathcal{G}_{S'} = (\mathcal{V}_{S'}, \mathcal{E}_{S'})$ is a subgraph of $\mathcal{G}_{S} = (\mathcal{V}_{S}, \mathcal{E}_{S})$ if $\mathcal{V}_{S'} \subseteq \mathcal{V}_{S}$ and $\mathcal{E}_{S'} \subseteq \mathcal{E}_{S}$.
Compared with sequence pattern mining, graph pattern mining is more general since it involves the structural topology and attribute information. For instance, as depicted in Figure~\ref{fig:multiple_session_graph}, \{\textit{cell phone}, \textit{tablet}, \textit{notebook}\} corresponds to a triangle. It may be challenging to discover the triangle from the temporal sequence, but it is a vital clue from the graph view. Thus, we stick on frequent graph patterns rather than sequence patterns.
\section{Methodology}

\label{sec:method}
We present the Frequent Attribute Pattern Augmented Transformer (FAPAT), a novel framework that captures user intents and item correlations. Our method is built upon session sequences and corresponding attribute graphs. Initially, we mine frequent attribute patterns from the attribute graphs to explore coarse-grained item correlations. These patterns are then used as memory to enhance the session encoder, which consists of graph-nested transformer layers.
Figure~\ref{fig:architecture} illustrates the overview.

\subsection{Frequent Attribute Pattern Acquisition}
\label{sec:pattern_mining}
In this subsection, we describe how to extract frequent patterns from training recommendation sessions. The aim of frequent pattern mining is to minimize the impact of random clicks in the global transition graph and avoid over-smoothing when learning multiplex attribute graphs.
To achieve this, we design a mining-filtering paradigm to ensure representativeness.

\subsubsection{Graph Pattern Mining}

Small graph patterns, also called motifs or graphlets, are valuable features in graph learning and property prediction~\cite{ShervashidzeVPMB09,KriegeJM20}, exhibiting strong statistical correlations between graph structures and node semantics. In this study, we focus on similar graphlet structures but extend to SBR scenarios. We collect patterns consisting of no more than four nodes (representing different attribute values), and further restrict them to those containing either a circle or a triangle to significantly reduce the number of candidates. We adopt gSpan~\cite{YanH02} to acquire undirected patterns from attribute session graphs and keep patterns belonging to one of twenty types shown in Figure~\ref{fig:architecture}.

\subsubsection{Loose Pattern Filtering}

\begin{figure*}[t]
  \centering
  \includegraphics[width=0.83\linewidth]{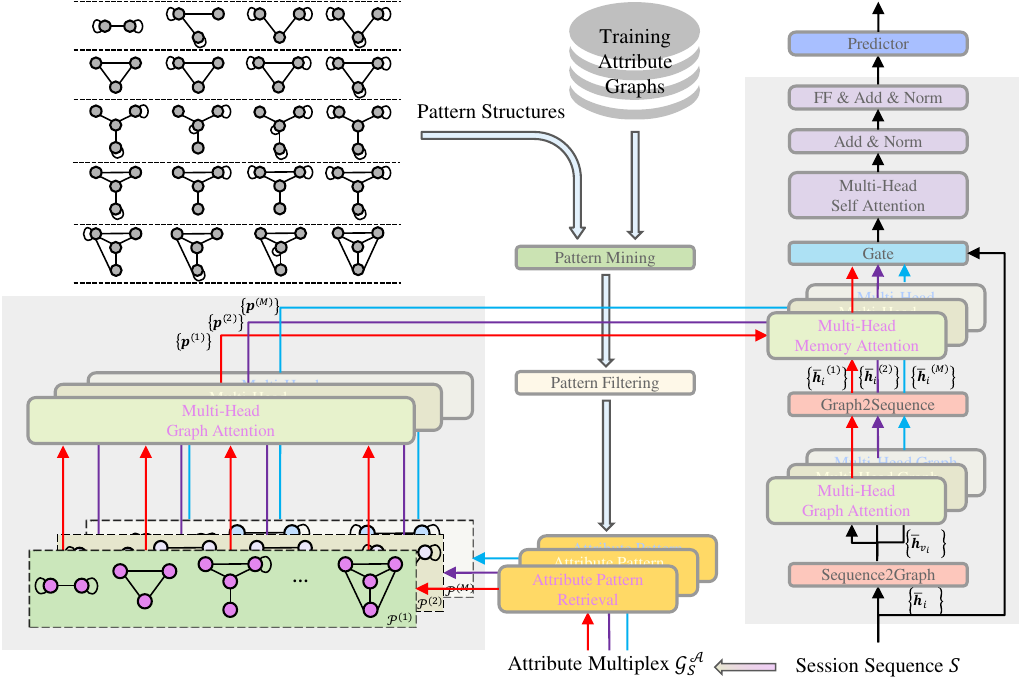}
  \captionof{figure}{An overview of FAPAT.}
  \label{fig:architecture}
  % \vspace{-0.2in}
\end{figure*}

However, complex patterns may contain smaller ones. For instance, the first pattern in Figure~\ref{fig:architecture} is a subgraph of the second. While each subgraph has an equal or higher frequency than its supergraph, such loose patterns do not convey much information. To eliminate them, we employ VF2~\cite{CordellaFSV04} subgraph isomorphism algorithm to filter. If a pattern $P'$ is a subgraph of another pattern $P$, then $P'$ is excluded from the pattern candidates. Finally, we retain compact and frequent patterns for session encoding.

\subsection{Intent-aware Sequence Encoding}
\label{sec:memory_augment}
To recommend items of high interest to users, we use a GAT-based encoder to learn pattern representations from the item side, which are then served as memory to augment session encoding.

\subsubsection{Relevant Graph Pattern Retrieval}
The input item sequence is converted to a multiplex session graph representing the transitional information of different item attributes, as depicted in Figure \ref{fig:same_session_graph}. To improve graph representations by utilizing frequent attribute subgraphs, we retrieve relevant patterns from those mined in \S \ref{sec:pattern_mining}. For the $m$-th attribute type $\mathcal{A}^{(m)} \ (1 \leq m \leq M)$, we denote an arbitrary subgraph mined from the previous step as $\mathcal{G}_{P}^{(m)} = (\mathcal{V}_{P}^{(m)}, \mathcal{E}_{P}^{(m)})$.
Then we retrieve at most $I$ subgraph patterns that have the most considerable Jaccard similarities to the transition graph $\mathcal{G}_{S}^{(m)}$ as Eq.~(\ref{eq:jaccard}).
This can be done within $\mathcal{O}(|\mathcal{V}_{S}^{(m)}| \times |\mathcal{V}_{P}^{(m)}|)$, but this is always linear because of $|\mathcal{V}_{P}^{(m)}| \leq 4$.
\begin{align}
    \text{Jaccard} \big( \mathcal{V}_{S}^{(m)},\mathcal{V}_{P}^{(m)} \big) = \frac{|\mathcal{V}_{S}^{(m)} \cap \mathcal{V}_{P}^{(m)}|}{|\mathcal{V}_{S}^{(m)}| + |\mathcal{V}_{P}^{(m)}| - |\mathcal{V}_{S}^{(m)} \cap \mathcal{V}_{P}^{(m)}|}. \label{eq:jaccard}
\end{align}

\subsubsection{Attribute Pattern Representation}

After retrieving the relevant patterns for the corresponding multiplex session graph, we encode them using multi-head relational graph attention for further memory augmentation. For a pattern $\mathcal{G}_{P}^{(m)}$ for the $m$-th attribute, we compute the attention weight for two arbitrary nodes $a_{i}^{(m)}$ and $a_{j}^{(m)}$ by:
\begin{align}
    \alpha_{ij}^{(m)} = \text{softmax}\big(\frac{\text{LeakyReLU}\big({\bm{r}_{ij}^{(m)}}^\top ( \bm{e}_{a_{j}^{(m)}} \circ \bm{e}_{a_{i}^{(m)}} )\big)}{\sum_{a_{k}^{(m)} \in \mathcal{N}(a_{i}^{(m)})} \text{LeakyReLU}\big({\bm{r}_{ik}^{(m)}}^\top (\bm{e}_{a_{k}^{(m)}} \circ \bm{e}_{a_{i}^{(m)}})\big)}\big),
    \label{eq:gat1}
\end{align}
where $\bm{e}_{a_{i}^{(m)}}$ denotes the embedding of $a_{i}^{(m)}$, $\circ$ indicates element-wise multiplication, $\mathcal{N}(a_{i}^{(m)})$ represents the neighbors of $a_{i}^{(m)}$ (including itself), and $\bm{r}_{ij}^{(m)}$ corresponds to the  relation-specific vector.
Then the representation of $ \overline{\bm{h}}_{a_{i}^{(m)}}$ for node $a_{i}^{(m)}$ in the pattern is aggregated by:
\begin{align}
   \overline{\bm{h}}_{a_{i}^{(m)}} = \sum_{a_{j}^{(m)} \in \mathcal{N}(a_{i}^{(m)})} \alpha_{ij}^{(m)} \bm{e}_{a_{i}^{(m)}}.
   \label{eq:gat2}
\end{align}
After computing the representation of each node in pattern subgraph $\mathcal{G}_{P}^{(m)}$, a pooling layer (e.g, average pooling) over all node representations is to aggregate the pattern presentation for $\mathcal{G}_{P}^{(m)}$:
\begin{align}
    \bm{p}^{(m)} = \text{Pool} \big( \{\overline{\bm{h}}_{a_{i}^{(m)}} | v_i \in \mathcal{G}_{P}^{(m)} \}\big) \label{eq:pattern_representation}
\end{align}

\subsubsection{Attribute Memory Augmentation}

Meanwhile, we also employ graph attention to obtain the native graph representations of the multiplex session graph $\mathcal{G}_{S}^{(m)}$. We apply this to each $m$-th attribute transition graph $\mathcal{G}_{S}^{(m)}$, computing node representations $a_{i}^{(m)}$ using Eq.~(\ref{eq:gat1}) and Eq.~(\ref{eq:gat2}) and denote the resulting node representation as $\overline{\bm{h}}_{v_i}^{(m)}$.
To compute the aggregated attribute representation of each node $v_i$ in the original session graph $\mathcal{G}_{S}$, we combine its different representations from all attribute transition graphs by:
\begin{align}
    \overline{\bm{h}}_{v_i} = \frac{1}{M} \sum_m \overline{\bm{h}}_{v_i}^{(m)} + \bm{e}_{v_i},
\end{align}
where $\bm{e}_{v_i}$ is the representation of $v_i$ by the item embedding lookup.

Suppose all retrieved attribute patterns associated with $\mathcal{A}^{(m)}$ are $\mathcal{P}^{(m)}$, whose representations from Eq.~(\ref{eq:pattern_representation}) are $\{\bm{p}^{(m)} | p^{(m)} \in \mathcal{P}^{(m)}\}$. To preserve temporal information when utilizing these patterns, we map graph representations $\{\overline{\bm{h}}_{v_i}^{(m)} | v_i \in \mathcal{V}_{S}\}$ to sequence representations $\{\overline{\bm{h}}_{i}^{(m)} | 1\leq i \leq L\}$ with length $L$. Next, we concatenate all pattern representations $\{{\bm{p}}^{(m)} | p^{(m)} \in \mathcal{P}^{(m)}\}$, all session representations $\{\overline{\bm{h}}_{i}^{(m)} | 1\leq i \leq L\}$, and two special embeddings $\bm{e}_{\text{CLS}}$ and $\bm{e}_{\text{MASK}}$ to separate the two parts and indicate the item to predict:
\begin{align}
    \Bigg[ \bigg[ \concat_{p^{(m)} \in \mathcal{P}^{(m)}}{{\bm{p}}^{(m)}} \bigg] \big\Vert \bm{e}_{\text{CLS}} \big\Vert \bigg[ \concat_{1\leq i \leq L}{\overline{\bm{h}}_{i}}^{(m)} \bigg] \big\Vert \bm{e}_{\text{MASK}}\Bigg]. \nonumber
\end{align}

\begin{figure}[t]
    \begin{minipage}{0.68\textwidth}
    \centering
    \includegraphics[width=\linewidth]{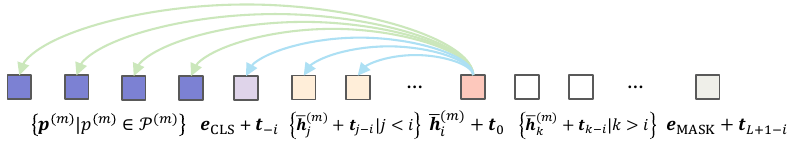}
    \end{minipage}
    \begin{minipage}{0.30\textwidth}
    \captionof{figure}{Schema of pattern augmented attention: relative position bias is added, and one item can only access memory and previous histories.}
    \label{fig:mem_attn}
    \end{minipage}
    \vspace{-0.2in}
\end{figure}

Then, we augment sequence encoding with memory, taking inspiration from TransformerXL~\cite{DaiYYCLS19} and Memorizing Transformer~\cite{WuRHS22}.
As shown in Figure~\ref{fig:mem_attn}, we introduce T5 relative position bias~\cite{RaffelSRLNMZLL20} to distinct short and long-term histories. Furthermore, we utilize unidirectional attention to focus solely on the user's current intent instead of global intent, which may become increasingly noisy over time.

\subsubsection{User Intent Aggregation}
To aggregate memory-augmented representations from $M$ different attributes and pattern sets, we employ gating mechanism:
\begin{align}
    \hat{\bm{h}}_{i} = \sum_{m=1}^{M} \beta_{i}^{(m)} \hat{\bm{h}}_{i}^{(m)} + \overline{\bm{h}}_{i}, 
    \text{s.t. }
    \beta_{i}^{(m)} = {\text{softmax}(\bm{W}_{\beta}^{(m)} \overline{\bm{h}}_{i}^{(m)} + \bm{b}_{\beta}^{(m)})_{}}, \label{eq:gate}
\end{align}
where $\bm{W}_{\beta}^{(m)} \in \mathbb{R}^{d \times d}$ and $\bm{b}_{\beta}^{(m)} \in \mathbb{R}^{d}$ are trainable parameters ($d$ is the dimension of hidden states), $\hat{\bm{h}}_{i}^{(m)}$ is the memory-augmented results, and $\beta_{i}^{(m)}$ controls the importance of attribute patterns.
While $\hat{\bm{h}}_{i}$ combines attribute information, it disregards long-range histories as $\overline{\bm{h}}_{i}^{(m)}$ solely accesses one-hop neighbors. Thus, we incorporate a transformer block to aggregate global session information:
\begin{align}
    \bm{H} = \text{Transformer} \big( \Bigg[\hat{\bm{h}}_{\text{CLS}} \big\Vert \bigg[ \concat_{1\leq i \leq L}{\hat{\bm{h}}_{i}} \bigg] \big\Vert \hat{\bm{h}}_{\text{MASK}}\Bigg] \big), \label{eq:item_rep}
\end{align}
where $\hat{\bm{h}}_{\text{CLS}}$ and $\hat{\bm{h}}_{\text{MASK}}$ are directly taken from Eq.~(\ref{eq:gate}) since they do not belong to the original session but still observe the memory patterns and relative temporal signals.

\subsection{Next-item Recommendation}
\label{sec:predictor}

Once the sequence representations are obtained, the next step is to predict the next item that may interest the user for clicking or purchasing.
We adopt the approach used in previous work~\cite{WuT0WXT19,WangWCLMQ20,XiaYYWC021} to concatenate additional reversed positional embeddings as follows:
\begin{align}
    \bm z_{i} = \text{tanh} \big( \bm W_{z} \big[ \bm{h}_{i} \parallel \bm t_{L-i+1} \big] + \bm b_{z} \big), 0 \leq i \leq L+1,
\end{align}
where $\bm{h}_{i}$ is the $i$-th item representation $\bm{H}[i]$ from Eq.~(\ref{eq:item_rep}), and $\bm{W}_{z} \in \mathbb{R}^{2d \times d}$ and $\bm{b}_{z} \in \mathbb{R}^{d}$ are trainable parameters. This concatenation of positional embeddings is intended to prioritize nearest intents over long-distance historical purposes. Figure~\ref{fig:multiple_session_graph} demonstrates this idea through two cases: the male user is more likely to be interested in Android cell phones since his last click was in that category, while the female user sticks to Apple products as she reviews iPhone and iPad once again.

We utilize the representation of the MASK to compute soft attention and then represent the session and the user's latest intent through a weighted average:
\begin{align}
    \quad \bm u = \sum_{i=1}^{L} \gamma_{i} \bm{h}_{i}, \text{s.t. }
    \gamma_{i} = \bm r_{\gamma}^\top \text{sigmoid}(\bm W_{\gamma} \bm z_{i} + \bm z_{\text{mask}} + \bm b_{\gamma}),
\end{align}
where $\bm{W}_{\gamma} \in \mathbb{R}^{d \times d}$ and $\bm{b}_{\gamma} \in \mathbb{R}^{d}$ are trainable parameters.
We compute the prediction of the next item using a similarity-based approach rather than a linear layer.
This approach is similar to the optimization of Bayesian Personalized Ranking (BPR)~\cite{RendleFGS09}.
Alternatively, we can train the model using cross-entropy minimization.
Finally, the prediction is obtained by $\hat{y} = \argmax_{v}(\bm u^\top \bm e_{v})$.
\section{Experiment}
\label{sec:experiment}

\subsection{Setup}
{\flushleft \textit{Datasets}.}
We first evaluate our method on public benchmarks.
\textit{diginetica}
contains the browser logs and anonymized transactions, \textit{Tmall}
collects anonymous users' shopping logs on the Tmall online website.
We also acquire sessions from the browse and purchase logs from our E-commerce platform.
We target at \textit{beauty}, \textit{books}, and \textit{electronics} and gather 20-minute interactions within last successful purchases into one session after removing long-tail items.
Appendix~\ref{appedix:data} provides more details.

{\flushleft \textit{Baselines}.}
We compare our method with seven sequence-based baselines
({FPMC}~\cite{RendleFS10},
{GRU4Rec}~\cite{HidasiKBT15},
{NARM}~\cite{LiRCRLM17},
{STAMP}~\cite{LiuZMZ18},
{CSRM}~\cite{WangRMCMR19},
{S3-Rec}~\cite{ZhouWZZWZWW20},
and {M2TRec}~\cite{ShalabyOAKC22})
and six graph-based baselines
({SR-GNN}~\cite{WuT0WXT19},
{GC-SAN}~\cite{XuZLSXZFZ19},
{S2-DHCN}~\cite{XiaYYWC021},
{GCE-GNN}~\cite{WangWCLMQ20},
{LESSR}~\cite{ChenW20},
 and {MSGIFSR}~\cite{Guo0SZWBZ22}).
Model details are given in Appendix~\ref{appendix:baseline}.
Each model is aligned with the official code implementation.

{\flushleft \textit{Evaluation}.}
We evaluate SBRs as a ranking problem and employ Hits@$K$, NDCG@$K$, MRR@$K$ as standard metrics.
Hits@$K$ measures the percentage of ranks up to and including $K$, while NDCG@$K$ assigns higher scores to hits at the top of the list. MRR@$K$ is the average of reciprocal ranks, with ranks above $K$ assigned 0. For public benchmarking, we report the average performance across seeds $\{2020, \cdots, 2024\}$, while for industrial data, we use a fixed seed (see Appendix~\ref{appendix:experiment} for more details).

\subsection{Next-item Prediction Evaluation}

\begin{table*}[t]
    \sisetup{detect-weight=true,detect-inline-weight=math}
    \centering
    \scriptsize
    \setlength\tabcolsep{2pt}
    \captionof{table}{Performance evaluation for next-item prediction, where standard deviations are enclosed in brackets.
The best and second-best results are respectively highlighted in bold and underlined.
Methods that use attributes are marked with \ddag, and * indicates the $p$-value $<$ 0.0001 in t-test.}
\label{tab:next_item_pub}
\vspace{-0.1in}
\begin{tabular}{l|ccc|ccc|ccc|ccc}
        \toprule
        \multicolumn{1}{c|}{\multirow{2}{*}{Model}} & 
        \multicolumn{6}{c|}{diginetica} & \multicolumn{6}{c}{Tmall}\\
        & \multicolumn{1}{c}{\tiny{Hits@10}} & \multicolumn{1}{c}{\tiny{NDCG@10}} & \multicolumn{1}{c|}{\tiny{MRR@10}}
        & \multicolumn{1}{c}{\tiny{Hits@20}} & \multicolumn{1}{c}{\tiny{NDCG@20}} & \multicolumn{1}{c|}{\tiny{MRR@20}}
        & \multicolumn{1}{c}{\tiny{Hits@10}} & \multicolumn{1}{c}{\tiny{NDCG@10}} & \multicolumn{1}{c|}{\tiny{MRR@10}}
        & \multicolumn{1}{c}{\tiny{Hits@20}} & \multicolumn{1}{c}{\tiny{NDCG@20}} & \multicolumn{1}{c}{\tiny{MRR@20}}
        \\
        \midrule
FPMC & 31.57* & 17.40* & 13.08* & 43.19* & 20.33* & 13.88* & 13.71* & 9.02* & 7.56* & 16.44* & 9.71* & 7.74* \\
GRU4Rec & \underline{36.77}* & 20.71* & 15.80* & \underline{49.68}* & \underline{23.97}* & 16.70* & 18.82* & 12.28* & {10.25*}* & 22.68* & 13.25* & 10.51* \\
NARM & 35.98* & 20.18* & 15.36* & 48.89* & 23.44* & 16.26* & 22.74* & 15.46* & 13.19* & 26.73* & 16.47* & 13.47* \\
STAMP & 33.59* & 18.89* & 14.41* & 45.87* & 22.00* & 15.26* & 24.32* & 16.55* & 14.12* & 28.40* & 17.58* & 14.41* \\
CSRM & 33.97* & 19.43* & 14.98* & 45.83* & 22.42* & 15.80* & 25.13* & 18.56* & 16.48* & 27.94* & 19.27* & 16.68* \\
S3-Rec\ddag & 33.48* & 18.58* & 14.04* & 45.97* & 21.74* & 14.90* & 18.24* & 12.30* & 10.46* & 22.31* & 13.32* & 10.74* \\
M2TRec\ddag & 29.67* & 16.30* & 12.23* & 41.23* & 19.22* & 13.02* & 11.42* & 7.56* & 6.36* & 13.75* & 8.15* & 6.52* \\
\hline
SR-GNN & 35.21* & 19.68* & 14.94* & 47.99* & 22.90* & 15.82* & 18.21* & 12.11* & 10.20* & 21.34* & 12.91* & 10.42* \\
GC-SAN & 35.25* & 19.72* & 14.97* & 47.87* & 22.90* & 15.85* & 19.29* & 12.80* & 10.78* & 23.18* & 13.78* & 11.05* \\
S2-DHCN & 30.76* & 17.04* & 12.86* & 42.39* & 19.98* & 13.66* & 22.00* & 13.36* & 10.68* & 27.23* & 14.69* & 11.05* \\
GCE-GNN & 36.32* & \underline{20.77}* & \underline{16.02}* & 48.67* & 23.89* & \underline{16.87}* & \underline{28.33}* & \underline{20.01}* & \underline{17.32}* & \underline{30.24}* & \underline{20.50}* & \underline{17.45}* \\
LESSR & 33.68* & 18.71* & 14.14* & 46.23* & 21.88* & 15.01* & 20.99* & 14.64* & 12.13* & 25.92* & 13.96* & 10.50* \\
MSGIFSR & 34.74* & 19.43* & 14.76* & 46.23* & 21.88* & 15.01* & 23.18* & 15.19* & 12.69* & 27.78* & 16.35* & 13.01* \\
\hline
FAPAT\ddag & \textbf{37.42} & \textbf{21.31} & \textbf{16.39} & \textbf{50.41} & \textbf{24.59} & \textbf{17.29} & \textbf{32.45} & \textbf{22.02} & \textbf{18.72} & \textbf{36.18} & \textbf{22.97} & \textbf{18.99} \\
\textit{Improv.} & \textit{3.03\%} & \textit{2.60\%} & \textit{2.31\%} & \textit{1.46\%} & \textit{2.59\%} & \textit{2.49\%} & \textit{14.19\%} & \textit{10.04\%} & \textit{8.08\%} & \textit{19.64\%} & \textit{12.05\%} & \textit{8.83\%}
\\
        \bottomrule
    \end{tabular}
    \vspace{-0.1in}
\end{table*}

\begin{table*}[t]
    \sisetup{detect-weight=true,detect-inline-weight=math}
    \centering
    \scriptsize
    \setlength\tabcolsep{2pt}
    \captionof{table}{Performance evaluation for next-item prediction on our 100 million industrial data.}
\label{tab:next_item_industry}
\vspace{-0.1in}
\begin{tabular}{l|cc|cc|cc|cc|cc|cc}
        \toprule
        \multicolumn{1}{c|}{\multirow{3}{*}{Model}} & 
        \multicolumn{4}{c|}{Beauty} & \multicolumn{4}{c|}{Books} & \multicolumn{4}{c}{Electronics} \\
        & \multicolumn{1}{c}{\tiny{Hits@10}} & \multicolumn{1}{c|}{\tiny{NDCG@10}}
        & \multicolumn{1}{c}{\tiny{Hits@20}} & \multicolumn{1}{c|}{\tiny{NDCG@20}}
        & \multicolumn{1}{c}{\tiny{Hits@10}} & \multicolumn{1}{c|}{\tiny{NDCG@10}}
        & \multicolumn{1}{c}{\tiny{Hits@20}} & \multicolumn{1}{c|}{\tiny{NDCG@20}}
        & \multicolumn{1}{c}{\tiny{Hits@10}} & \multicolumn{1}{c|}{\tiny{NDCG@10}}
        & \multicolumn{1}{c}{\tiny{Hits@20}} & \multicolumn{1}{c}{\tiny{NDCG@20}}
        \\
        \midrule
CSRM & 89.74 & 75.28 & 92.61 & 76.01 & \underline{78.69} & 56.70 & \underline{82.88} & 57.77 & 62.28 & 44.35 & 67.47 & 45.67 \\
S3-Rec\ddag & 89.64 & \underline{75.56} & 92.53 & \underline{76.30} & 75.00 & \underline{58.54} & 79.45 & \underline{59.67} & 74.36 & \underline{56.03} & 79.63 & \underline{57.37} \\
M2TRec\ddag & 80.13 & 65.97 & 83.66 & 66.87 & 32.56 & 22.58 & 35.39 & 25.70 & 57.32 & 44.84 & 61.70 & 45.95 \\
\hline
SR-GNN & 88.69 & 70.42 & 91.74 & 71.20 & 66.55 & 47.55 & 69.77 & 48.37 & \underline{74.86} & 54.30 & \underline{79.66} & 55.52 \\
GCE-GNN & 89.34 & 73.15 & 91.29 & 73.65 & 77.61 & 57.60 & 80.03 & 58.22 & 72.93 & 53.74 & 78.49 & 55.15 \\
LESSR & 89.95 & 71.29 & \underline{92.98} & 72.06 & 73.72 & 53.86 & 82.31 & 54.77 & 72.91 & 50.46 & 78.78 & 51.96 \\
MSGIFSR & \underline{90.18} & 73.62 & 92.50 & 74.21 & 72.93 & 52.23 & 76.33 & 53.09 & 73.56 & 53.83 & 77.45 & 54.73 \\
\hline
FAPAT\ddag & \textbf{92.72} & \textbf{76.29} & \textbf{94.10} & \textbf{76.87} & \textbf{81.62} & \textbf{61.08} & \textbf{85.12} & \textbf{61.97} & \textbf{78.36} & \textbf{56.81} & \textbf{82.81} & \textbf{57.94} \\
\textit{Improv.} & \textit{2.82\%} & \textit{0.97\%} & \textit{1.20\%} & \textit{0.75\%} & \textit{3.72\%} & \textit{4.34\%} & \textit{2.70\%} & \textit{3.85\%} & \textit{4.68\%} & \textit{1.39\%} & \textit{3.95\%} & \textit{0.99\%} \\
        \bottomrule
    \end{tabular}
    \vspace{-0.15in}
\end{table*}

\paragraph{Experimental Results.}
We first compare our models with selected baselines on two public datasets in Table~\ref{tab:next_item_pub}. Overall, there is no huge difference between sequence and graph models when the session historical information is limited (e.g., in \textit{diginetica}). However, we observe a performance boost with the history attention mechanism and session topology. FPMC that utilizes first-order Markov chains and matrix factorization is the worst. This reveals the difference between traditional recommendation and session-based recommendation.
In contrast, RNN-based methods (GRU4Rec, NARM, STAMP, and CSRM) show better generalizability, along with the benefits of attention and memory. But we do not see further gains from pretraining and multi-task learning in S3-Rec and M2TRec. On the other hand, GNN-based algorithms for local sessions also achieve comparable results even with the absence of temporal signals. Moreover, the heterogeneous graphs of MSGIFSR beat the shortcut graphs from LESSR, indicating over-smoothing and potential noise. Although explicit global collaborative graphs enhance GCE-GNN, the implicit collaborative information from neurons and patterns in FAPAT easily surpasses other baselines significantly.
At the same time, we evaluate algorithms on our 100 million industrial data. Table~\ref{tab:next_item_industry} demonstrates that most sequence models become unstable except S3-Rec. It employs incomplete data to predict masked items and attributes during pretraining and identifies contextualized collaborations by determining whether two incomplete sequences belong to the same session. But the multi-tasking learning for predicting attributes makes M2TRec hard to recommend items. While most GNN methods are competitive, S3-Rec and FAPAT are still better.

\paragraph{Over-smoothing Relief.} 
We examine different graph topologies in baselines and FAPAT by comparing the graph density of different methods:
the density of local session graphs (in SR-GNN), global collaborative filtering graphs(in GCE-GNN), shortcut graphs (in LESSR), heterogeneous graphs (in MSGIFSR), attribute patterns (in FAPAT) on E-commerce data are 3.658, 114.910, 25.447, 2.598, and 1.117, respectively.
The density decrease significantly relieves the over-smoothing.
Detailed analyses and expectations are provided in Appendix~\ref{appendix:density}.

\subsection{Ablation Study}

\setcounter{figure}{{\value{figure}-1}}

\begin{figure}[t]
    \centering
    \begin{subfigure}[b]{.50\textwidth}
        \centering
        % \vspace{-0.15in}
        \includegraphics[height=5.0cm]{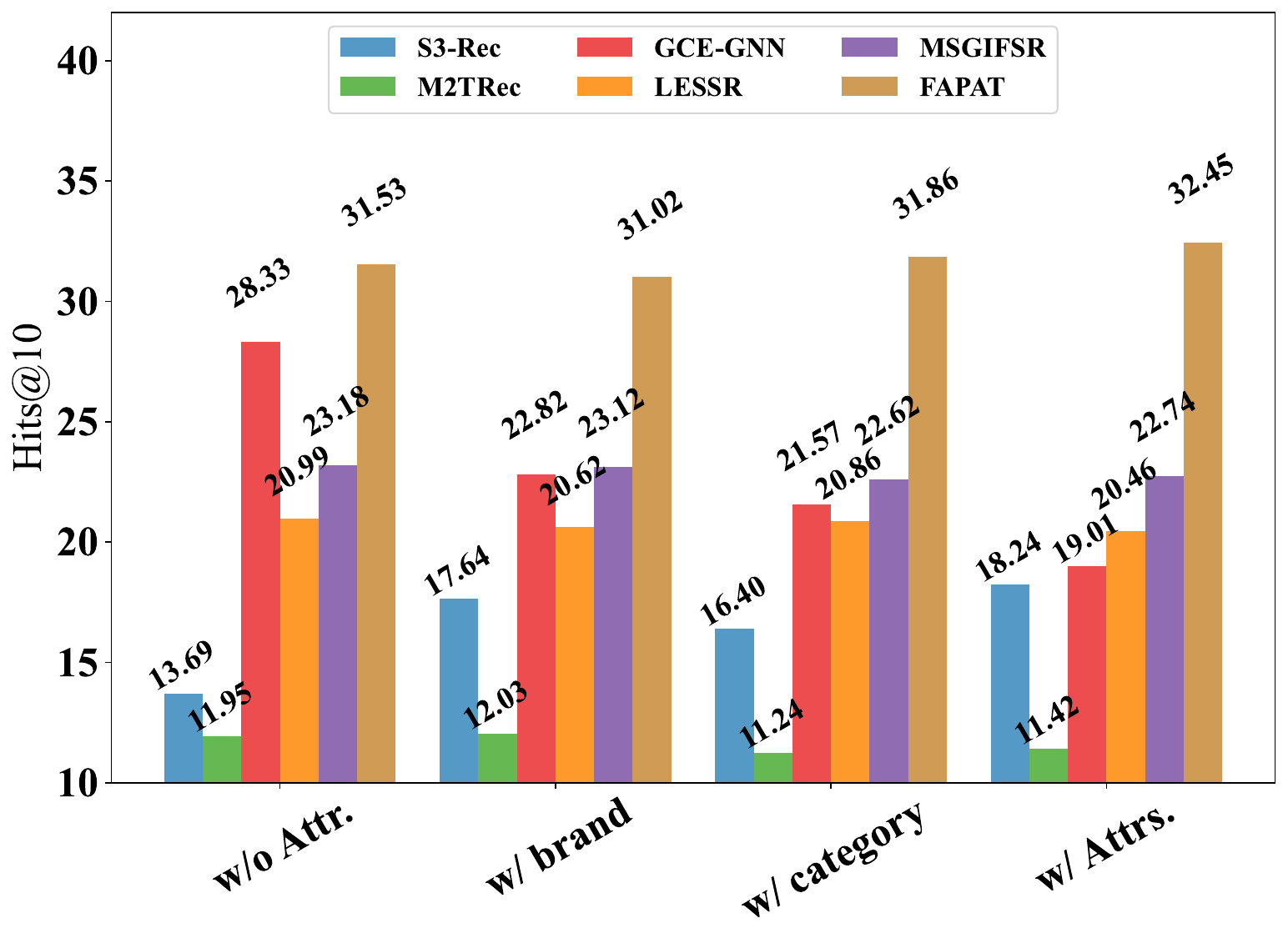}
        % \includegraphics[width=4.0cm]{figures/line2_svg-tex.pdf}
        % \vspace{-0.25in}
        \caption{Tmall}
        \label{fig:tmall_attr}
    \end{subfigure}
    \begin{subfigure}[b]{.46\textwidth}
        \centering
        % \vspace{-0.15in}
        \includegraphics[height=5.0cm]{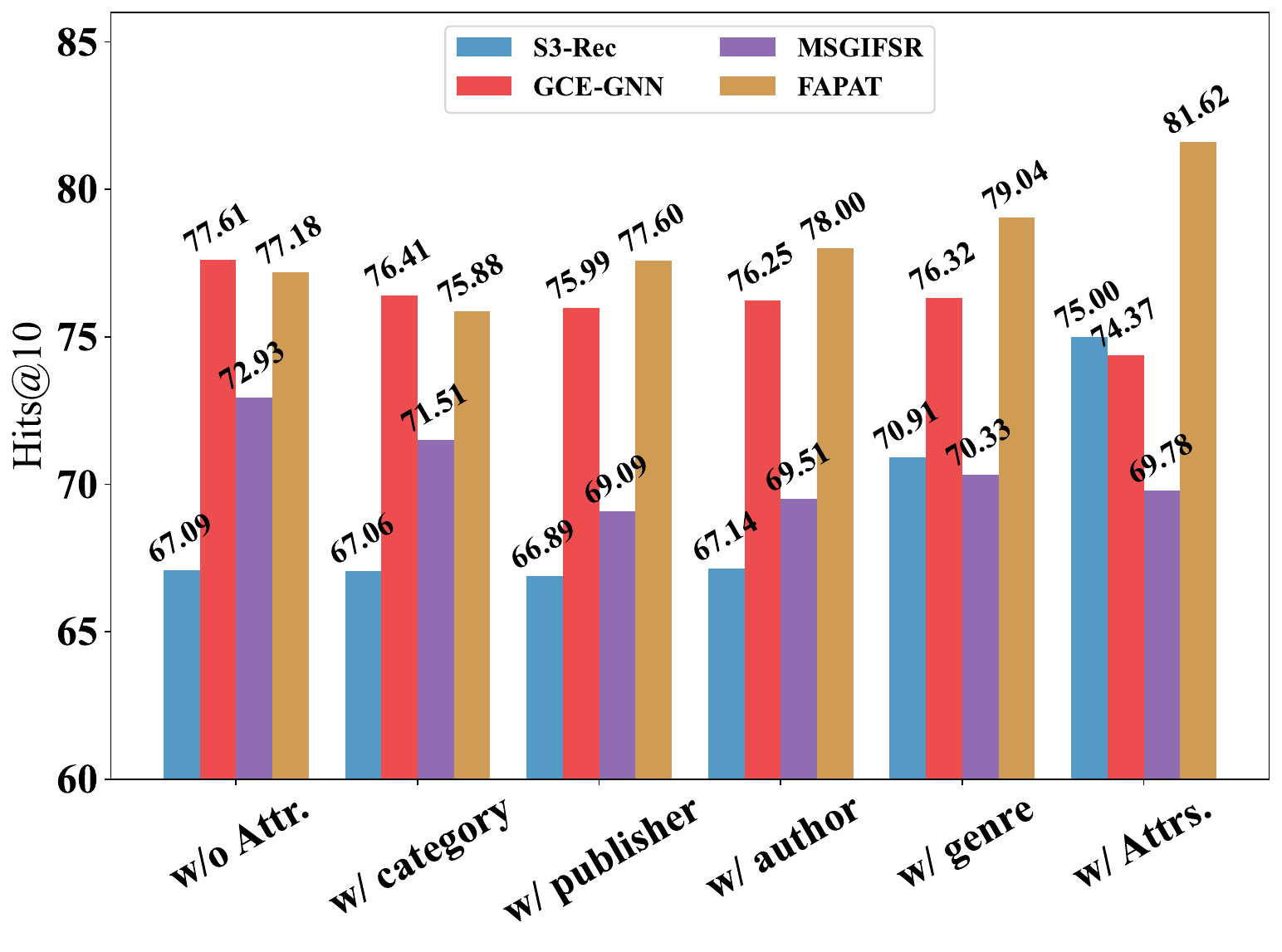}
        % \includegraphics[width=4.0cm]{figures/line5_svg-tex.pdf}
        % \vspace{-0.25in}
        \caption{Books}
        \label{fig:book_attr}
    \end{subfigure}
    \captionof{figure}{Effects of different attribute settings.}
    \label{fig:attr}
\end{figure}

\paragraph{Attribute Pattern Augmentation.}

To evaluate the effect of attribute patterns, we conduct experiments on variants with single attributes or without any attribute.
The same soft attention strategy from Eq.~(\ref{eq:gate}) is employed to fuse attribute embeddings for competitive baselines.
Results in Figure~\ref{fig:attr} show that attribute pattern augmentation is more stable than attribute soft attention.
FAPAT benefits from graph-nested attention, where the graph attention aligns the hidden space, and the memory attention captures item correlations and user intents.
But attribute embeddings may have side effects on optimization in baselines, especially in graph neural networks.
We also discover that not all attributes have a positive impact.
Comparing among attribute pattern numbers, attribute patterns with significant frequencies (slightly lower than or similar to the item number) can have adverse effects.

\paragraph{Graph-nested Attention.}
The graph-nested attention is one of our contributions, distinguishing it from GraphFormer~\cite{YangLXLLASSX21}.
Unlike GraphFormer, our graph attention is integrated inside the blocks, which allows for direct benefit from the broader attention in the following self-attention via backpropagation.
To ensure fairness, we replace the encoding module of FAPAT with GraphFormer and vanilla Transformer.
Results in Table~\ref{tab:attn} demonstrate the advantages of our proposed graph-nested attention.
Our experiments also show that even a simple vanilla Transformer can outperform previous state-of-the-art models by a significant margin, indicating the importance of emphasizing temporal information in SBRs and the appropriateness of attention for capturing long-distance dependencies.

\begin{table}[t]
    \sisetup{detect-weight=true,detect-inline-weight=math}
    \centering
    \footnotesize
    \setlength\tabcolsep{2.5pt}
    \captionof{table}{Results of encoder comparison, where * indicates the p-value < 0.05 in t-test.}
    \label{tab:attn}
    \begin{tabular}{l|ll|ll|ll|ll}
        \toprule
        \multicolumn{1}{c|}{\multirow{2}{*}{Encoder}} & \multicolumn{2}{c|}{\textit{diginetica}} & \multicolumn{2}{c|}{\textit{Tmall}} & \multicolumn{2}{c|}{\textit{Beauty}} & \multicolumn{2}{c}{\textit{Books}} \\
        & \multicolumn{1}{c}{{Hits@10}} & \multicolumn{1}{c|}{{MRR@10}} & \multicolumn{1}{c}{{Hits@10}} & \multicolumn{1}{c|}{{MRR@10}} & \multicolumn{1}{c}{{Hits@10}} & \multicolumn{1}{c|}{{MRR@10}} & \multicolumn{1}{c}{{Hits@10}} & \multicolumn{1}{c}{{MRR@10}} \\
        \midrule
        FAPAT w/o Attr. & \hspace{0.6em}\textbf{36.82} & \hspace{0.6em}16.29 & \hspace{0.6em}\textbf{31.53} & \hspace{0.6em}\textbf{18.86} & \hspace{0.6em}\textbf{88.70} & \hspace{0.6em}\textbf{67.11} & \hspace{0.6em}\textbf{77.18} & \hspace{0.6em}\textbf{49.66} \\
        GraphFormer & \hspace{0.6em}36.05* & \hspace{0.6em}16.17 & \hspace{0.6em}30.05* & \hspace{0.6em}18.58 & \hspace{0.6em}88.48 & \hspace{0.6em}65.65 & \hspace{0.6em}77.03 & \hspace{0.6em}47.86 \\
        Transformer & \hspace{0.6em}36.30* & \hspace{0.6em}16.02* & \hspace{0.6em}28.83* & \hspace{0.6em}18.30* & \hspace{0.6em}88.10 & \hspace{0.6em}65.90 & \hspace{0.6em}74.67 & \hspace{0.6em}46.54 \\
        \bottomrule
    \end{tabular}
\end{table}

\begin{table}[t]
    \sisetup{detect-weight=true,detect-inline-weight=math}
    \centering
    \footnotesize
    \setlength\tabcolsep{4pt}
    \captionof{table}{Attribute estimation evaluation.}
    \label{tab:attr}
    \begin{tabular}{l|ll|ll|ll|ll}
        \toprule
        \multicolumn{1}{c|}{\multirow{2}{*}{Model}} & \multicolumn{2}{c|}{\textit{diginetica}} & \multicolumn{2}{c|}{\textit{Tmall}} & \multicolumn{2}{c|}{\textit{Beauty}} & \multicolumn{2}{c}{\textit{Books}} \\
        & \multicolumn{1}{c}{{Hits@10}} & \multicolumn{1}{c|}{{MRR@10}} & \multicolumn{1}{c}{{Hits@10}} & \multicolumn{1}{c|}{{MRR@10}} & \multicolumn{1}{c}{{Hits@10}} & \multicolumn{1}{c|}{{MRR@10}} &  \multicolumn{1}{c}{{Hits@10}} & \multicolumn{1}{c}{{MRR@10}} \\
        \midrule
        CSRM & \hspace{0.6em}89.87* & \hspace{0.6em}87.91* & \hspace{0.6em}48.03* & \hspace{0.6em}34.97* & \hspace{0.6em}94.82 & \hspace{0.6em}83.57 & \hspace{0.6em}88.93 & \hspace{0.6em}73.31 \\
        M2TRec\ddag & \hspace{0.6em}\textbf{94.10} & \hspace{0.6em}\textbf{89.76} & \hspace{0.6em}\underline{59.44} & \hspace{0.6em}39.72 & \hspace{0.6em}\underline{95.93} & \hspace{0.6em}84.95 & \hspace{0.6em}83.80 & \hspace{0.6em}71.22 \\
        S3-Rec\ddag & \hspace{0.6em}89.66* & \hspace{0.6em}\underline{88.42}* & \hspace{0.6em}44.95* & \hspace{0.6em}32.60* & \hspace{0.6em}95.35 & \hspace{0.6em}84.79 & \hspace{0.6em}88.11 & \hspace{0.6em}75.63 \\
        \hline
        GCE-GNN & \hspace{0.6em}89.89* & \hspace{0.6em}87.96* & \hspace{0.6em}55.88* & \hspace{0.6em}38.57* & \hspace{0.6em}95.44 & \hspace{0.6em}84.16 & \hspace{0.6em}\underline{89.78} & \hspace{0.6em}\underline{75.68} \\
        LESSR & \hspace{0.6em}88.60* & \hspace{0.6em}86.24* & \hspace{0.6em}49.43* & \hspace{0.6em}32.61* & \hspace{0.6em}95.02 & \hspace{0.6em}81.56 & \hspace{0.6em}89.57 & \hspace{0.6em}73.73 \\
        MSGIFSR & \hspace{0.6em}89.48* & \hspace{0.6em}87.19* & \hspace{0.6em}50.09* & \hspace{0.6em}33.95* & \hspace{0.6em}95.79 & \hspace{0.6em}\underline{85.42} & \hspace{0.6em}86.67 & \hspace{0.6em}71.99 \\
        \hline
        FAPAT\ddag & \hspace{0.6em}\underline{89.99}* & \hspace{0.6em}88.23* & \hspace{0.6em}\textbf{59.49} & \hspace{0.6em}\textbf{40.56} & \hspace{0.6em}\textbf{95.94} & \hspace{0.6em}\textbf{86.87} & \hspace{0.6em}\textbf{90.77} & \hspace{0.6em}\textbf{76.83} \\
        \ w/o Attr.  & \hspace{0.6em}89.35* & \hspace{0.6em}87.50* & \hspace{0.6em}58.22* & \hspace{0.6em}\underline{40.08} & \hspace{0.6em}95.07 & \hspace{0.6em}81.65 & \hspace{0.6em}89.29 & \hspace{0.6em}74.12 \\
        \bottomrule
    \end{tabular}
\end{table}

\setcounter{figure}{{\value{figure}-1}}
\begin{figure}[t]
\centering
    \begin{subfigure}[b]{.32\textwidth}
        \centering
        % \vspace{-0.15in}
        \includegraphics[height=4.8cm]{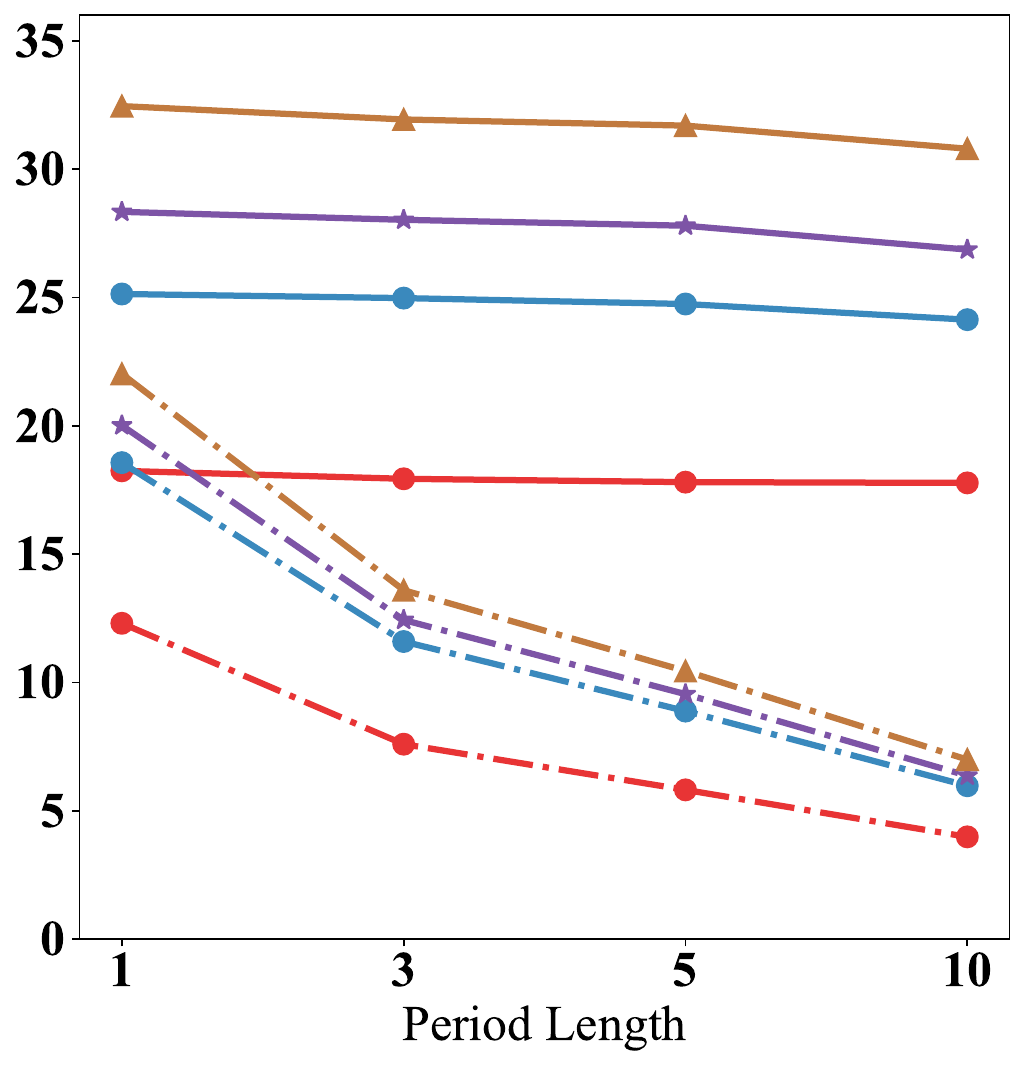}
        % \vspace{-0.25in}
        \caption{Tmall}
        \label{fig:tmall_period}
    \end{subfigure}
    \ \ 
    \begin{subfigure}[b]{.32\textwidth}
        \centering
        % \vspace{-0.15in}
        \includegraphics[height=4.8cm]{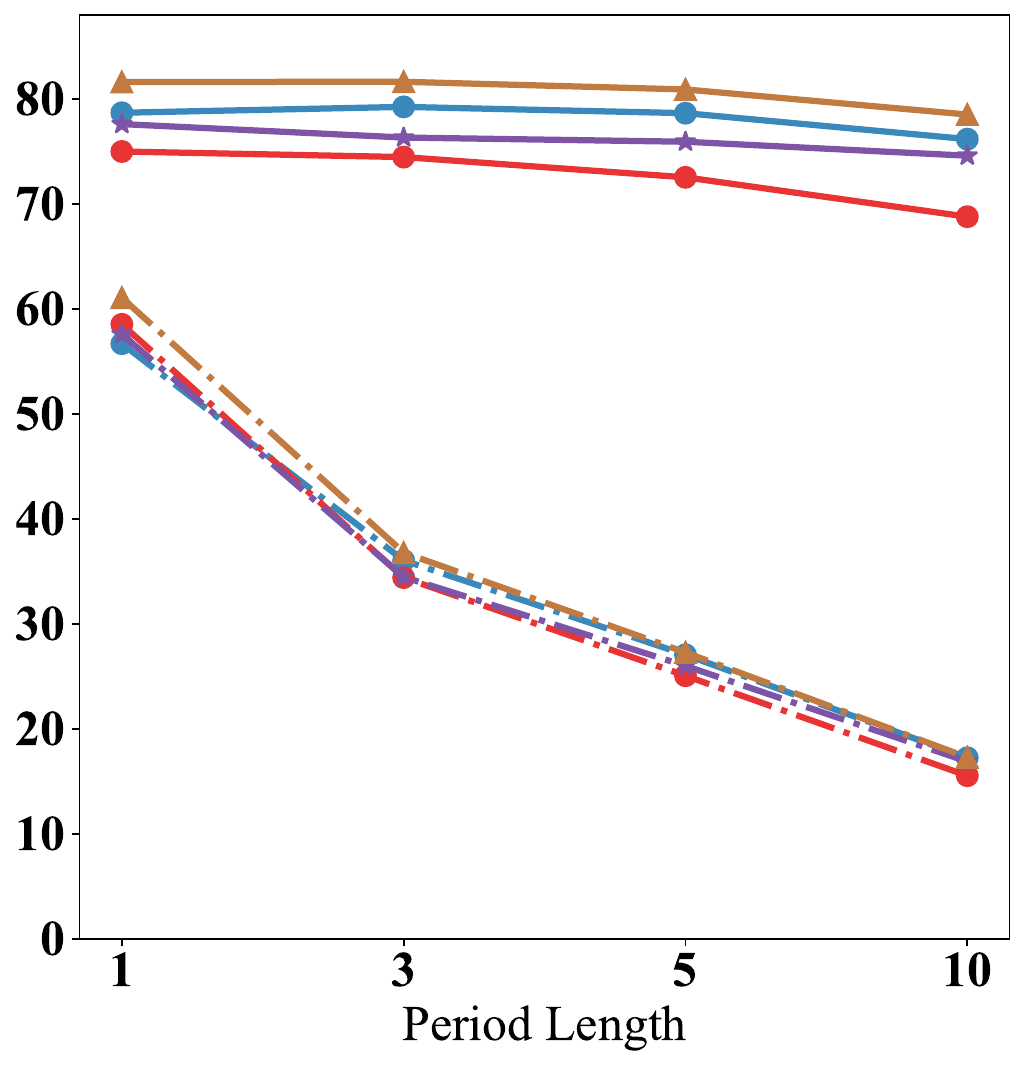}
        % \vspace{-0.25in}
        \caption{Books}
        \label{fig:book_period}
    \end{subfigure}
    \ \ 
    \begin{subfigure}[b]{.32\textwidth}
        \centering
        % \vspace{-0.15in}
        \includegraphics[height=4.8cm]{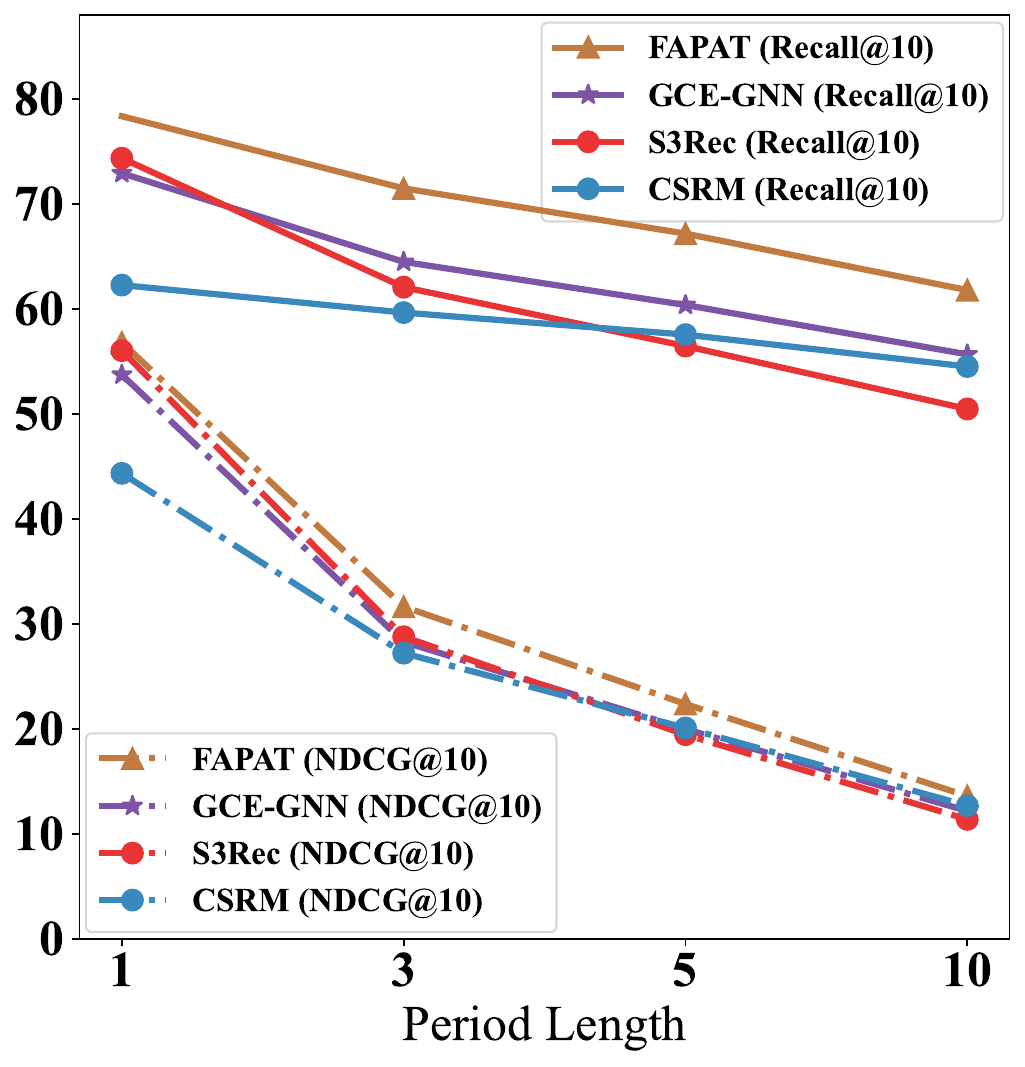}
        % \vspace{-0.25in}
        \caption{Electronics}
        \label{fig:elec_period}
    \end{subfigure}
    \captionof{figure}{Period-item recommendation evaluation.}
    \vspace{-0.2in}
    \label{fig:period}
\end{figure}

\subsection{Intent Capture Inspection}
\label{sec:intent_capture}
\paragraph{Attribute Estimation.}

Beyond item predictions, we also estimate the awareness of user intents from the product attribute side.
We do not require models (except M2TRec) to predict attributes but to retrieve attributes from predicted items instead.
We consider it a successful estimation if the top-ranked items have the same attribute value as the ground truth.
M2TRec performs well on public data with multi-task learning but struggles on industrial E-commerce data with four attributes, as shown in Table~\ref{tab:attr}.
Pretraining in S3-Rec is not helpful due to catastrophic forgetting.
GNN-based models perform similarly, except on \textit{Tmall}, where the data are too sparse so that global collaborative information assists.
On the contrary, FAPAT achieves robust predictions across four datasets.
Even when session data are sufficient, frequent patterns remain effective.

\paragraph{Period-item Recommendation.}
In addition to single-step evaluation, we revisit recommenders.
A sound and robust SRB system must understand user intents in deep and foresee the possible consistency and the potential change.
Autoregressive settings pose challenges for GNNs due to short click histories and error accumulation.
Therefore, we evaluate period recommendations like search engines by comparing the top-10 predicted items with the next 3/5/10 clicks.
Figure~\ref{fig:period} demonstrates period-recommendation performance.
Sequence models offer steady results in Recall, indicating that temporal information is one of the prerequisites to analyzing users' latest intents.
However, pretraining may hinder models' ability to generalize over long periods, resulting in a severe decline in NDCG.
Our FAPAT achieves the best performance in all metrics, indicating the effectiveness of attribute graphlets in capturing deep user intents.
\section{Conclusion}
\label{sec:conclusion}

% This paper proposes a novel framework called FAPAT, which effectively leverages attribute graph patterns to augment the anonymous sequence encoding for session-based recommendations.
% Compared with the GNN-based methods, frequent attribute graphlets can significantly reduce the noise and the computational cost.
% By combining graph attention for patterns and local session graphs, memory attention along with session sequences, and Transformer for global information aggregation, our proposed sequence encoder can better preserve the temporal signals and capture the user's latest intents.
% Experimental results on public benchmarks and industrial data clearly illustrate the efficiency of our algorithm.
% And the extensive ablation study, attribute estimations, and period-item recommendations also firmly support our claims.
Our paper introduces FAPAT, a novel framework that leverages attribute graph patterns to augment anonymous sequence encoding for session-based recommendations. Compared to other GNN-based methods, frequent attribute graphlets can reduce noise and topology densities for enhancing user intent capture. Our sequence encoder can better preserve temporal signals and forecast the user's latest intents. Experimental results clearly illustrate the effectiveness, and extensive ablation studies and intent capture inspections provide additional support. We discuss limitations in Appendix~\ref{appendix:limitation}.
One of the future works is to improve ranking priority by combining pretraining and pattern augmentation.

\section*{Acknowledgments}
The authors of this paper were supported by the NSFC Fund (U20B2053) from the NSFC of China, the RIF (R6020-19 and R6021-20) and the GRF (16211520 and 16205322) from RGC of Hong Kong. We also thank the support from the UGC Research Matching Grants (RMGS20EG01-D, RMGS20CR11, RMGS20CR12, RMGS20EG19, RMGS20EG21, RMGS23CR05, RMGS23EG08).

\clearpage
\newpage
\bibliographystyle{plainnat}
\bibliography{neurips_2023}

\begin{thebibliography}{47}
\providecommand{\natexlab}[1]{#1}
\providecommand{\url}[1]{\texttt{#1}}
\expandafter\ifx\csname urlstyle\endcsname\relax
  \providecommand{\doi}[1]{doi: #1}\else
  \providecommand{\doi}{doi: \begingroup \urlstyle{rm}\Url}\fi

\bibitem[Chen and Wong(2020)]{ChenW20}
Tianwen Chen and Raymond~Chi{-}Wing Wong.
\newblock Handling information loss of graph neural networks for session-based recommendation.
\newblock In \emph{{SIGKDD}}, pages 1172--1180, 2020.

\bibitem[Chen et~al.(2009)Chen, Kuo, Wu, and Tang]{ChenKWT09}
Yen{-}Liang Chen, Mi{-}Hao Kuo, Shin{-}yi Wu, and Kwei Tang.
\newblock Discovering recency, frequency, and monetary {(RFM)} sequential patterns from customers' purchasing data.
\newblock \emph{Electronic Commerce Research and Applications}, 8\penalty0 (5):\penalty0 241--251, 2009.

\bibitem[Cordella et~al.(2004)Cordella, Foggia, Sansone, and Vento]{CordellaFSV04}
Luigi~P. Cordella, Pasquale Foggia, Carlo Sansone, and Mario Vento.
\newblock A (sub)graph isomorphism algorithm for matching large graphs.
\newblock \emph{{TPAMI}}, 26\penalty0 (10):\penalty0 1367--1372, 2004.

\bibitem[Cui et~al.(2022)Cui, Ma, Zhou, Zhou, and Yang]{cui2022m6}
Zeyu Cui, Jianxin Ma, Chang Zhou, Jingren Zhou, and Hongxia Yang.
\newblock M6-rec: Generative pretrained language models are open-ended recommender systems.
\newblock \emph{arXiv preprint arXiv:2205.08084}, 2022.

\bibitem[Dai et~al.(2019)Dai, Yang, Yang, Carbonell, Le, and Salakhutdinov]{DaiYYCLS19}
Zihang Dai, Zhilin Yang, Yiming Yang, Jaime~G. Carbonell, Quoc~Viet Le, and Ruslan Salakhutdinov.
\newblock Transformer-xl: Attentive language models beyond a fixed-length context.
\newblock In \emph{{ACL}}, pages 2978--2988, 2019.

\bibitem[Geng et~al.(2022)Geng, Liu, Fu, Ge, and Zhang]{Geng0FGZ22}
Shijie Geng, Shuchang Liu, Zuohui Fu, Yingqiang Ge, and Yongfeng Zhang.
\newblock Recommendation as language processing {(RLP):} {A} unified pretrain, personalized prompt {\&} predict paradigm {(P5)}.
\newblock In \emph{{RecSys}}, pages 299--315, 2022.

\bibitem[Guo et~al.(2022)Guo, Yang, Song, Zhang, Wang, Bai, and Zhang]{Guo0SZWBZ22}
Jiayan Guo, Yaming Yang, Xiangchen Song, Yuan Zhang, Yujing Wang, Jing Bai, and Yan Zhang.
\newblock Learning multi-granularity consecutive user intent unit for session-based recommendation.
\newblock In \emph{{WSDM}}, pages 343--352, 2022.

\bibitem[Hidasi et~al.(2016)Hidasi, Karatzoglou, Baltrunas, and Tikk]{HidasiKBT15}
Bal{\'{a}}zs Hidasi, Alexandros Karatzoglou, Linas Baltrunas, and Domonkos Tikk.
\newblock Session-based recommendations with recurrent neural networks.
\newblock In \emph{{ICLR}}, 2016.

\bibitem[Hou et~al.(2022)Hou, Hu, Zhang, and Zhao]{HouHZZ22}
Yupeng Hou, Binbin Hu, Zhiqiang Zhang, and Wayne~Xin Zhao.
\newblock {CORE:} simple and effective session-based recommendation within consistent representation space.
\newblock In \emph{{SIGIR}}, pages 1796--1801. {ACM}, 2022.

\bibitem[Hu et~al.(2020)Hu, Dong, Wang, and Sun]{HuDWS20}
Ziniu Hu, Yuxiao Dong, Kuansan Wang, and Yizhou Sun.
\newblock Heterogeneous graph transformer.
\newblock In \emph{{WWW}}, pages 2704--2710, 2020.

\bibitem[Huang et~al.(2023)Huang, Gao, Li, Yang, Song, Zhang, Zhu, Jiang, Chang, and Yin]{huang2023ccgen}
Jie Huang, Yifan Gao, Zheng Li, Jingfeng Yang, Yangqiu Song, Chao Zhang, Zining Zhu, Haoming Jiang, Kevin Chen-Chuan Chang, and Bing Yin.
\newblock Ccgen: Explainable complementary concept generation in e-commerce.
\newblock \emph{arXiv preprint arXiv:2305.11480}, 2023.

\bibitem[Jiang et~al.(2022)Jiang, Cao, Li, Luo, Tang, Yin, Zhang, Goutam, and Yin]{jiang2022short}
Haoming Jiang, Tianyu Cao, Zheng Li, Chen Luo, Xianfeng Tang, Qingyu Yin, Danqing Zhang, Rahul Goutam, and Bing Yin.
\newblock Short text pre-training with extended token classification for e-commerce query understanding.
\newblock \emph{arXiv preprint arXiv:2210.03915}, 2022.

\bibitem[Jin et~al.(2023)Jin, Mao, Li, Jiang, Luo, Wen, Han, Lu, Wang, Li, Li, Cheng, Goutam, Zhang, Subbian, Wang, Sun, Tang, Yin, and Tang]{jin2023amazon}
Wei Jin, Haitao Mao, Zheng Li, Haoming Jiang, Chen Luo, Hongzhi Wen, Haoyu Han, Hanqing Lu, Zhengyang Wang, Ruirui Li, Zhen Li, Monica~Xiao Cheng, Rahul Goutam, Haiyang Zhang, Karthik Subbian, Suhang Wang, Yizhou Sun, Jiliang Tang, Bing Yin, and Xianfeng Tang.
\newblock Amazon-m2: A multilingual multi-locale shopping session dataset for recommendation and text generation.
\newblock In \emph{NeurIPS}, 2023.

\bibitem[Kingma and Ba(2015)]{KingmaB14}
Diederik~P. Kingma and Jimmy Ba.
\newblock Adam: {A} method for stochastic optimization.
\newblock In \emph{{ICLR}}, 2015.

\bibitem[Kriege et~al.(2020)Kriege, Johansson, and Morris]{KriegeJM20}
Nils~M. Kriege, Fredrik~D. Johansson, and Christopher Morris.
\newblock A survey on graph kernels.
\newblock \emph{Appl. Netw. Sci.}, 5\penalty0 (1):\penalty0 6, 2020.

\bibitem[Li et~al.(2017)Li, Ren, Chen, Ren, Lian, and Ma]{LiRCRLM17}
Jing Li, Pengjie Ren, Zhumin Chen, Zhaochun Ren, Tao Lian, and Jun Ma.
\newblock Neural attentive session-based recommendation.
\newblock In \emph{{CIKM}}, pages 1419--1428, 2017.

\bibitem[Lin et~al.(2022)Lin, Tian, Hou, and Zhao]{LinTHZ22}
Zihan Lin, Changxin Tian, Yupeng Hou, and Wayne~Xin Zhao.
\newblock Improving graph collaborative filtering with neighborhood-enriched contrastive learning.
\newblock In \emph{{WWW}}, pages 2320--2329, 2022.

\bibitem[Liu et~al.(2021)Liu, Cheng, Zhu, Gao, and Nie]{LiuCZGN21}
Fan Liu, Zhiyong Cheng, Lei Zhu, Zan Gao, and Liqiang Nie.
\newblock Interest-aware message-passing {GCN} for recommendation.
\newblock In \emph{{WWW}}, pages 1296--1305, 2021.

\bibitem[Liu et~al.(2018)Liu, Zeng, Mokhosi, and Zhang]{LiuZMZ18}
Qiao Liu, Yifu Zeng, Refuoe Mokhosi, and Haibin Zhang.
\newblock {STAMP:} short-term attention/memory priority model for session-based recommendation.
\newblock In \emph{{SIGKDD}}, pages 1831--1839, 2018.

\bibitem[Liu and Song(2022)]{LiuS22}
Xin Liu and Yangqiu Song.
\newblock Graph convolutional networks with dual message passing for subgraph isomorphism counting and matching.
\newblock In \emph{{AAAI}}, pages 7594--7602, 2022.

\bibitem[Liu et~al.(2020)Liu, Pan, He, Song, Jiang, and Shang]{LiuPHSJS20}
Xin Liu, Haojie Pan, Mutian He, Yangqiu Song, Xin Jiang, and Lifeng Shang.
\newblock Neural subgraph isomorphism counting.
\newblock In \emph{{SIGKDD}}, pages 1959--1969. {ACM}, 2020.

\bibitem[Liu et~al.(2022)Liu, Cheng, Song, and Jiang]{LiuCSJ22}
Xin Liu, Jiayang Cheng, Yangqiu Song, and Xin Jiang.
\newblock Boosting graph structure learning with dummy nodes.
\newblock In \emph{{ICML}}, volume 162, pages 13704--13716. {PMLR}, 2022.

\bibitem[Luo et~al.(2022)Luo, Headden, Avudaiappan, Jiang, Cao, Yin, Gao, Li, Goutam, Zhang, and Yin]{chen22query}
Chen Luo, William Headden, Neela Avudaiappan, Haoming Jiang, Tianyu Cao, Qingyu Yin, Yifan Gao, Zheng Li, Rahul Goutam, Haiyang Zhang, and Bing Yin.
\newblock Query attribute recommendation at amazon search.
\newblock In \emph{RecSys}, page 506–508, 2022.

\bibitem[Qiu et~al.(2019)Qiu, Li, Huang, and Yin]{QiuLHY19}
Ruihong Qiu, Jingjing Li, Zi~Huang, and Hongzhi Yin.
\newblock Rethinking the item order in session-based recommendation with graph neural networks.
\newblock In \emph{{CIKM}}, pages 579--588, 2019.

\bibitem[Raffel et~al.(2020)Raffel, Shazeer, Roberts, Lee, Narang, Matena, Zhou, Li, and Liu]{RaffelSRLNMZLL20}
Colin Raffel, Noam Shazeer, Adam Roberts, Katherine Lee, Sharan Narang, Michael Matena, Yanqi Zhou, Wei Li, and Peter~J. Liu.
\newblock Exploring the limits of transfer learning with a unified text-to-text transformer.
\newblock \emph{Journal of Machine Learning Research}, 21:\penalty0 140:1--140:67, 2020.

\bibitem[Rendle et~al.(2009)Rendle, Freudenthaler, Gantner, and Schmidt{-}Thieme]{RendleFGS09}
Steffen Rendle, Christoph Freudenthaler, Zeno Gantner, and Lars Schmidt{-}Thieme.
\newblock {BPR:} bayesian personalized ranking from implicit feedback.
\newblock In \emph{{UAI}}, pages 452--461, 2009.

\bibitem[Rendle et~al.(2010)Rendle, Freudenthaler, and Schmidt{-}Thieme]{RendleFS10}
Steffen Rendle, Christoph Freudenthaler, and Lars Schmidt{-}Thieme.
\newblock Factorizing personalized markov chains for next-basket recommendation.
\newblock In \emph{{WWW}}, pages 811--820, 2010.

\bibitem[Shalaby et~al.(2022)Shalaby, Oh, Afsharinejad, Kumar, and Cui]{ShalabyOAKC22}
Walid Shalaby, Sejoon Oh, Amir Afsharinejad, Srijan Kumar, and Xiquan Cui.
\newblock M2trec: Metadata-aware multi-task transformer for large-scale and cold-start free session-based recommendations.
\newblock In \emph{{RecSys}}, pages 573--578, 2022.

\bibitem[Shervashidze et~al.(2009)Shervashidze, Vishwanathan, Petri, Mehlhorn, and Borgwardt]{ShervashidzeVPMB09}
Nino Shervashidze, S.~V.~N. Vishwanathan, Tobias Petri, Kurt Mehlhorn, and Karsten~M. Borgwardt.
\newblock Efficient graphlet kernels for large graph comparison.
\newblock In \emph{{AISTATS}}, volume~5, pages 488--495, 2009.

\bibitem[Wan et~al.(2023)Wan, Liu, Wang, Qiu, Li, Guo, Chen, and Wang]{wan2023spatio}
Zhongwei Wan, Xin Liu, Benyou Wang, Jiezhong Qiu, Boyu Li, Ting Guo, Guangyong Chen, and Yang Wang.
\newblock Spatio-temporal contrastive learning enhanced gnns for session-based recommendation.
\newblock \emph{Transactions on Information Systems}, 2023.

\bibitem[Wang et~al.(2019)Wang, Ren, Mei, Chen, Ma, and de~Rijke]{WangRMCMR19}
Meirui Wang, Pengjie Ren, Lei Mei, Zhumin Chen, Jun Ma, and Maarten de~Rijke.
\newblock A collaborative session-based recommendation approach with parallel memory modules.
\newblock In \emph{{SIGIR}}, pages 345--354, 2019.

\bibitem[Wang et~al.(2022{\natexlab{a}})Wang, Li, Sun, Liu, Li, Yin, and Abdelzaher]{wang22nips}
Ruijie Wang, Zheng Li, Dachun Sun, Shengzhong Liu, Jinning Li, Bing Yin, and Tarek Abdelzaher.
\newblock Learning to sample and aggregate: Few-shot reasoning over temporal knowledge graphs.
\newblock In \emph{NeurIPS}, pages 16863--16876, 2022{\natexlab{a}}.

\bibitem[Wang et~al.(2022{\natexlab{b}})Wang, Zhang, Hu, Zhang, Wang, and Aggarwal]{WangZ0ZW022}
Shoujin Wang, Qi~Zhang, Liang Hu, Xiuzhen Zhang, Yan Wang, and Charu Aggarwal.
\newblock Sequential/session-based recommendations: Challenges, approaches, applications and opportunities.
\newblock In \emph{{SIGIR}}, pages 3425--3428, 2022{\natexlab{b}}.

\bibitem[Wang et~al.(2020)Wang, Wei, Cong, Li, Mao, and Qiu]{WangWCLMQ20}
Ziyang Wang, Wei Wei, Gao Cong, Xiao{-}Li Li, Xianling Mao, and Minghui Qiu.
\newblock Global context enhanced graph neural networks for session-based recommendation.
\newblock In \emph{{SIGIR}}, pages 169--178, 2020.

\bibitem[Wu et~al.(2019)Wu, Tang, Zhu, Wang, Xie, and Tan]{WuT0WXT19}
Shu Wu, Yuyuan Tang, Yanqiao Zhu, Liang Wang, Xing Xie, and Tieniu Tan.
\newblock Session-based recommendation with graph neural networks.
\newblock In \emph{{AAAI}}, pages 346--353, 2019.

\bibitem[Wu et~al.(2022)Wu, Rabe, Hutchins, and Szegedy]{WuRHS22}
Yuhuai Wu, Markus~N. Rabe, DeLesley Hutchins, and Christian Szegedy.
\newblock Memorizing transformers.
\newblock In \emph{{ICLR}}, 2022.

\bibitem[Xia et~al.(2021)Xia, Yin, Yu, Wang, Cui, and Zhang]{XiaYYWC021}
Xin Xia, Hongzhi Yin, Junliang Yu, Qinyong Wang, Lizhen Cui, and Xiangliang Zhang.
\newblock Self-supervised hypergraph convolutional networks for session-based recommendation.
\newblock In \emph{{AAAI}}, pages 4503--4511, 2021.

\bibitem[Xu et~al.(2019{\natexlab{a}})Xu, Zhao, Liu, Sheng, Xu, Zhuang, Fang, and Zhou]{XuZLSXZFZ19}
Chengfeng Xu, Pengpeng Zhao, Yanchi Liu, Victor~S. Sheng, Jiajie Xu, Fuzhen Zhuang, Junhua Fang, and Xiaofang Zhou.
\newblock Graph contextualized self-attention network for session-based recommendation.
\newblock In \emph{{IJCAI}}, pages 3940--3946, 2019{\natexlab{a}}.

\bibitem[Xu et~al.(2019{\natexlab{b}})Xu, Hu, Leskovec, and Jegelka]{XuHLJ19}
Keyulu Xu, Weihua Hu, Jure Leskovec, and Stefanie Jegelka.
\newblock How powerful are graph neural networks?
\newblock In \emph{{ICLR}}, 2019{\natexlab{b}}.

\bibitem[Yan and Han(2002)]{YanH02}
Xifeng Yan and Jiawei Han.
\newblock gspan: Graph-based substructure pattern mining.
\newblock In \emph{{ICDM}}, pages 721--724, 2002.

\bibitem[Yang et~al.(2021)Yang, Liu, Xiao, Li, Lian, Agrawal, Singh, Sun, and Xie]{YangLXLLASSX21}
Junhan Yang, Zheng Liu, Shitao Xiao, Chaozhuo Li, Defu Lian, Sanjay Agrawal, Amit Singh, Guangzhong Sun, and Xing Xie.
\newblock Graphformers: Gnn-nested transformers for representation learning on textual graph.
\newblock In \emph{{NeurIPS}}, pages 28798--28810, 2021.

\bibitem[Yap et~al.(2012)Yap, Li, and Yu]{YapLY12}
Ghim{-}Eng Yap, Xiaoli Li, and Philip~S. Yu.
\newblock Effective next-items recommendation via personalized sequential pattern mining.
\newblock In \emph{{DASFAA}}, volume 7239, pages 48--64, 2012.

\bibitem[Yu et~al.(2023)Yu, Wang, Liu, Bai, Song, Li, Gao, Cao, and Yin]{yu-etal-2023-folkscope}
Changlong Yu, Weiqi Wang, Xin Liu, Jiaxin Bai, Yangqiu Song, Zheng Li, Yifan Gao, Tianyu Cao, and Bing Yin.
\newblock {F}olk{S}cope: Intention knowledge graph construction for {E}-commerce commonsense discovery.
\newblock In \emph{Finding of ACL}, pages 1173--1191, 2023.

\bibitem[Zhang et~al.(2021)Zhang, Li, Cao, Luo, Wu, Lu, Song, Yin, Zhao, and Yang]{zhang21queaco}
Danqing Zhang, Zheng Li, Tianyu Cao, Chen Luo, Tony Wu, Hanqing Lu, Yiwei Song, Bing Yin, Tuo Zhao, and Qiang Yang.
\newblock Queaco: Borrowing treasures from weakly-labeled behavior data for query attribute value extraction.
\newblock In \emph{CIKM}, page 4362–4372, 2021.

\bibitem[Zhang et~al.(2023)Zhang, Guo, Li, Xie, Kim, Zhang, Xie, Wang, and Kim]{abs-2206-12781}
Peiyan Zhang, Jiayan Guo, Chaozhuo Li, Yueqi Xie, Jaeboum Kim, Yan Zhang, Xing Xie, Haohan Wang, and Sunghun Kim.
\newblock Efficiently leveraging multi-level user intent for session-based recommendation via atten-mixer network.
\newblock In \emph{{WSDM}}, 2023.

\bibitem[Zhao and Akoglu(2020)]{ZhaoA20}
Lingxiao Zhao and Leman Akoglu.
\newblock Pairnorm: Tackling oversmoothing in gnns.
\newblock In \emph{{ICLR}}, 2020.

\bibitem[Zhou et~al.(2020)Zhou, Wang, Zhao, Zhu, Wang, Zhang, Wang, and Wen]{ZhouWZZWZWW20}
Kun Zhou, Hui Wang, Wayne~Xin Zhao, Yutao Zhu, Sirui Wang, Fuzheng Zhang, Zhongyuan Wang, and Ji{-}Rong Wen.
\newblock S3-rec: Self-supervised learning for sequential recommendation with mutual information maximization.
\newblock In \emph{{CIKM}}, pages 1893--1902, 2020.

\end{thebibliography}

%%%%%%%%%%%%%%%%%%%%%%%%%%%%%%%%%%%%%%%%%%%%%%%%%%%%%%%%%%%%

\clearpage
\appendix
\section{Limitations}
\label{appendix:limitation}
There are three limitations to our current proposed method and evaluation.
First, our method separately processes and retrieves patterns for each attribute type.
We do not merge all attributes in a candidate pool because we aim for our method to easily generalize to real recommendation systems with hundreds of attribute types and category hierarchies.
The current implementation supports adding a new attribute type to a model as long as its embeddings align with the embeddings of other attributes.
Second, we conducted experiments based on "clean" session data.
Most E-commerce platforms do not have truly clean data on product attributes, so attribute data, in general, is very sparse and full of invalid values.
We performed human-centric attribute regularization to drop products without valid attribute values, which may create a gap compared to a real industrial system.
Third, the evaluation does not consider the same products with different identifiers.
Therefore, evaluating results (especially MRR) cannot accurately reflect the performance.
To better reflect the real performance with error tolerance, a larger K is suggested.
The current comparison is still fair for all algorithms, and we address this synonym problem in attribute estimation in \S~\ref{sec:intent_capture}, where we merge attribute values based on semantics and syntax.

\section{Transition Graph Density}
\label{appendix:density}

\begin{figure}[t]
    \centering
    \includegraphics[width=\linewidth]{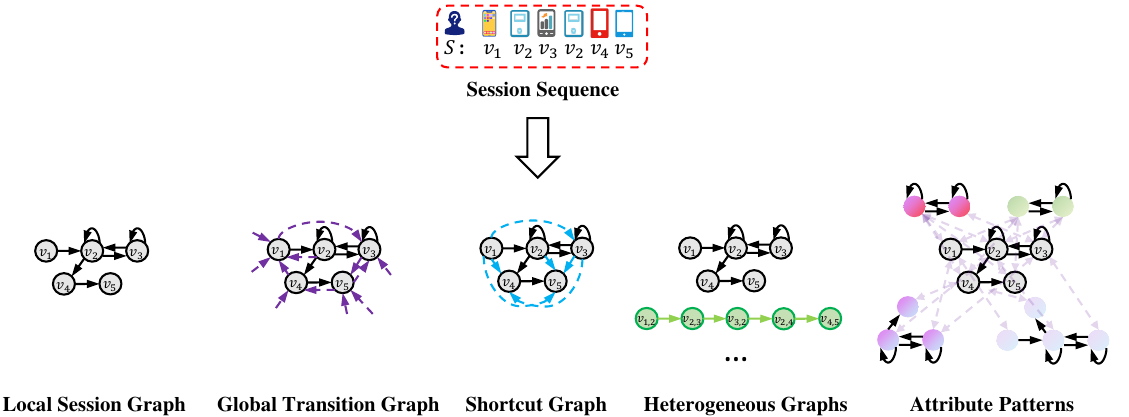}
    \captionof{figure}{
    Different granularities of graphs derived from a session, where the violet edges, blue edges, and green edges correspond to the global collaborations, shortcuts, and higher-level heterogeneous connections, respectively.}
    \label{fig:graph_density}
    \vspace{-0.2in}
\end{figure}

The graph structure is crucial for neural networks to capture explicit transitions and implicit connections.
A local session records the history of a user's clicks or purchases, which is usually sparse.
In contrast, the global collaborative graph could be extremely dense because each pair of items may have a potential connection.
From this perspective, the density indicates the explicit information provided from session data.
On the other hand, different graph topologies and densities also present different focuses and challenges.
The sparse local transition graph emphasizes current intents, while the global collaboration indicates broader interests and revenues.
Graph neural networks excel at capturing local features, but a large number of neighbors can overshadow important connections with less significant ones.
Considering that previous methods have focused on different granularities individually, we summarize them in Figure~\ref{fig:graph_density} and compare them in terms of optimization interpretations\footnote{Graphs are typically considered undirected in practical algorithms.}.
\begin{itemize}[leftmargin=*]
\item \textbf{Local} session graphs correspond to item transitions within a session, where edges are created between two consecutively clicked/purchased items. The density is usually sparse (slightly greater than 1.0), allowing exploration and global collaboration to be learned through model parameters instead of explicit connections. Therefore, generalizing to unseen click patterns becomes challenging.
\item \textbf{Global} transition graphs record all collaborations. In a real industrial system, the density is usually beyond one hundred or even one thousand. Ideally, any session can benefit from this global collaborative information, including multi-hop connections. However, optimizing graph neural networks to learn such topologies (due to oversmoothing) and building a large model for real-time inference (due to latency and streaming processing) pose challenges.
\item \textbf{Shortcut} graphs aim to avoid constructing global graphs and make session learning more efficient. They were proposed by LESSR~\cite{ChenW20} to address information loss in graph convolutions. Specifically, they allow latent items to be aware of all previous clicks, resembling shortcuts for multi-hop neighbors in the directed local session graph. However, they lack the extensive exploration capabilities of the global transition graph and suffer from oversmoothing issues due to dense connections.
\item \textbf{Heterogeneous} graphs strike a balance between shortcuts and local adjacency. Nodes with different numbers of items are categorized into different groups, and transition edges capture varying levels of spatial continuity. From a high-level perspective, this graph is sparser than the local session graph, resulting in faster convergence for optimization. However, the propagation of high-order information introduces additional processing costs and the risk of overfitting.
\item \textbf{Patterns}, especially attribute patterns, should be the most efficient features for recommendations in a large candidate item pool. Each pattern can be considered a higher-grained heterogeneous graph. However, pattern filtering can significantly eliminate noise influence, not to mention the benefits gained from offline indexing. Besides, the partial match of patterns can provide the intent information from other sessions, making the learning and prediction more reliable and steady.
\end{itemize}

\section{Experimental Data}
\label{appedix:data}

\subsection{Public Benchmarks}

We choose two public benchmarks for session-based recommendation evaluation:
\textit{diginetica}
\footnote{\url{https://competitions.codalab.org/competitions/11161}} 
is CIKM Cup 2016 that 
contains the browser logs and anonymized transactions; \textit{Tmall}
\footnote{\url{https://tianchi.aliyun.com/dataset/dataDetail?dataId=42}} 
comes from a competition in IJCAI-15 which 
collects anonymous users’ shopping logs on the Tmall online website.
We acquire attributes from the original data and drop items without attributes or with invalid values. Therefore,  the performance of baselines may not be exactly same as the reported numbers in the original papers.

\subsection{E-commerce Data Collection}
\label{appendix:data_collection}
We collect E-commerce data from our log systems in two months.
We follow the same procedure to clean and process session data in \textit{beauty}, \textit{books}, and \textit{electronics} domains\footnote{The sampled data scales and distributions are different in real systems due to out-of-domain items filtering.}:
\begin{enumerate}[label=\Roman*,leftmargin=*]
    \item We focus on successful purchases so that we only keep sessions ending with ``purchase'' actions. \\
    \item To make sure previous clicks can reflect the purchase intent, we drop actions 20 minutes ago. \\
    \item We filter out items with missing attributes (i.e., books without publishers, authors, or genre, and electronics without colors and brands). \\
    \item We adopt the 20-core setting to finalize the item sets, in which items appear on at least 20 different days. \\
    \item Only sessions whose length is no greater than 50 are preserved. \\
    \item We retrieve item attributes in our attribute databases.
    \item For GNN models that requires the global transition graph from training data, we maintain 12 neighbors based on the co-occurrence, which is consistent with GCE-GNN~\cite{WangWCLMQ20}.
\end{enumerate}

\subsection{Data Split}
\label{appendix:data_split}
We follow previous settings that split training/validation/testing data based on timestamps.
For \textit{diginetica}, we gather the last 8-14 days as validation, the last 7 days as testing, and remaining as training.
For \textit{Tmall}, we use the last 101-200 seconds as validation, the last 100 seconds as testing, and remaining as training.
For our industrial E-commerce data (i.e., \textit{Beauty}, \textit{Books}, \textit{Electronics}), we select the last 6-10 days as validation, the last 5 days as testing, and remaining as training.

\subsection{Data Statistics}
\label{appendix:statistics}
Table~\ref{tab:statistics_time} summarizes the statistics of the experimental datasets based on timestamps. The density is calculated based on undirected graphs, which would be doubled during graph convolution in practice.
\textit{Local density}, as used in SR-GNN and GC-SAN, corresponds to the average density of local session graphs in E-commerce sessions. On the other hand, \textit{global density}, as used in GCE-GNN, refers to the density of the global collaborative graph obtained by connecting all adjacent items appearing in all sessions. \textit{Shortcut density}, as used in LESSR, is the density resulting from connecting all items in a single session as a complete graph. \textit{Heterogeneous density}, as used in MSGIFSR, refers to the average density of the heterogeneous graphs obtained by regarding the consecutive adjacent two nodes as a fine-grained intent unit. Lastly, \textit{pattern density}, as used in FAPAT, is the density of the acquired frequent and compact patterns.
From Table~\ref{tab:statistics_time}, it is evident that leveraging patterns is the most effective way of characterizing user intents because other graph topologies vary with data sources and scales, making it difficult to generalize and provide stable performance. 
Besides, patterns can be preprocessed as indicies to aid recommendations, making them more practical in industrial scenarios. 
Moreover, it is easy to update attribute patterns dynamically, whereas other graph structures are more closely coupled with input sessions and are more sensitive to tiny variations.

\begin{table*}[t]
    \centering
    \scriptsize
    \setlength\tabcolsep{4pt}
    \caption{Statistics of datasets based on timestamps.}
    \label{tab:statistics_time}
    \begin{tabular}{l|c|c|c|c|c}
        \toprule
         & \multicolumn{2}{c|}{Public} & \multicolumn{3}{c}{Industrial (E-commerce)} \\
         & diginetica & Tmall & Beauty & Books & Electronics \\
        \midrule
        \#User & 57,623 & 7,576 & 2.6 M & 3.2 M & 10.2 M \\
        \#Item & 43,074 & 39,768 & 39.2 K & 94.8 K & 244.7 K \\
        \#Click & 993,163 & 438,315 & 27.2 M & 38.8 M & 115.6 M \\
        Avg.
        Len. & 4.850 & 6.649 & 10.325 & 11.912 & 11.249 \\
        \hline
        \#Train & 630,789 & 303,181 & 19.6 M & 28.2 M & 84.1 M \\
        \#Valid & 78,708 & 33,735 & 2.4 M & 3.5 M & 10.5 M \\
        \#Test & 78,907 & 35,481 & 2.5 M & 3.8 M & 10.6 M \\
        \hline
        \#Attribute & \specialcell{category : 995} & \specialcell{category : 821 \\ brand : 4,304} &  \specialcell{category : 359 \\ color : 1,101 \\ brand : 4,359 \\ size : 1,883} & \specialcell{category : 18 \\ publisher : 2,751 \\ author : 27,651 \\ genre : 2,634} & \specialcell{type : 123 \\ category : 881 \\ color : 2,096 \\ brand : 24,196} \\
        \hline
        \#Pattern & \specialcell{category : 1,866} & \specialcell{category : 33,582 \\ brand : 2,497} & \specialcell{category : 970 \\ color : 4,059 \\ brand : 254 \\ size : 1,091} & \specialcell{category : 24 \\ publisher : 4,370 \\ author : 1,399 \\ genre : 12,535} & \specialcell{type : 9,289 \\ category : 13,991 \\ color : 146,402 \\ brand : 14,043} \\
        \hline
        Density & \specialcell{Local: 0.886 \\ Global: 11.329 \\ Shortcut: 2.512 \\ Heterogeneous: 0.543 \\ Pattern: 1.023} & \specialcell{Local: 1.249 \\ Global: 10.222 \\ Shortcut: 4.983 \\ Heterogeneous: 0.707 \\ Pattern: 1.165} & \specialcell{Local: 4.510 \\ Global: 70.504 \\ Shortcut: 29.827 \\ Heterogeneous: 3.412 \\ Pattern: 1.095} & \specialcell{Local: 3.554 \\ Global: 99.389 \\ Shortcut: 26.649 \\ Heterogeneous: 2.333 \\ Pattern: 1.085} & \specialcell{Local: 2.910 \\ Global: 128.041 \\ Shortcut: 19.865 \\ Heterogeneous: 2.049 \\ Pattern: 1.189} \\
        \bottomrule
    \end{tabular}
\end{table*}

\section{Baselines}
\label{appendix:baseline}
We compare our method with the following baselines: \\
\noindent \textit{Sequence-based methods}
\begin{itemize}[leftmargin=*]
    \item \textbf{FPMC}~\cite{RendleFS10} learns the representation of session via Markov-chain based methods.
    \item \textbf{GRU4Rec}~\cite{HidasiKBT15} is the first RNN-based approach that simulates the Markov Decision Process (MDP) but has a better generalization.
    \item \textbf{NARM}~\cite{LiRCRLM17} is a attention-based RNN model to learn session embeddings.
    \item \textbf{STAMP}~\cite{LiuZMZ18} adopts attention mechanism between the last item to previous histories to represent users' short-term interests.
    \item \textbf{CSRM}~\cite{WangRMCMR19} proposes to engage an inner memory encoder and external memory network to capture correlations between neighborhood sessions to enrich the collaborative representations.
    \item \textbf{S3-Rec}~\cite{ZhouWZZWZWW20} is the first pretrained SBR model that predicts items, attributes, and segments during the pretraining stage.
    \item \textbf{M2TRec}~\cite{ShalabyOAKC22} is a metadata-aware multi-task Transformer model. In the original paper, the authors ignore item embeddings. For a fair comparison, we also regard the item ids as one of metadata.
\end{itemize}
\noindent \textit{Graph-based methods}
\begin{itemize}[leftmargin=*]
    \item \textbf{SR-GNN}~\cite{WuT0WXT19} is the first GNN-based model for the SBR task, which transforms the session data into a direct unweighted graph and learns the representation of the item-transitions graph.
    \item \textbf{GC-SAN}~\cite{XuZLSXZFZ19} uses gated GNNs to extract local context information and then self-attention to obtain the global representation.
    \item \textbf{S2-DHCN}~\cite{XiaYYWC021} transforms the session data into hyper-graphs and line-graphs and encodes them via GCNs to enhance the session representations.
    \item \textbf{GCE-GNN}~\cite{WangWCLMQ20} aggregates two levels of item embeddings from session graphs and global graphs with soft attention.
    \item \textbf{LESSR}~\cite{ChenW20}  preserves the edge order and constructs shortcuts to encode sessions for GNNs.
    \item \textbf{MSGIFSR}~\cite{Guo0SZWBZ22} captures the user intents from multiple granularities to relieve the computational burden of long-dependency. In experiments, we search the best model from the level-1, level-2, and level-3 consecutive intent units.
\end{itemize}

\section{Experimental Settings}
\label{appendix:experiment}
We fix all embeddings and hidden dimensions as 100, and the batch size is searched among \{100, 200, 500\} for all methods.
We also choose the number of layers/iterations (if applicable) from the validation performance (e.g., MRR@10).
A learning scheduler with 10\% linear warmup and 90\% decay is associated with the Adam optimizer~\cite{KingmaB14}.
The initial learning rate is set as 1e-3, and the regularization weight is tuned among \{1e-4, 1e-5, 1e-6\}.
We seek the dropout probability between two modules from \{0.0, 0.2, 0.4\}, but fix the attention dropout rate as 0.2.
The number of attention heads is empirically set as 4.
We follow the setting of GCE-GNN that the maximum one-hop neighbor number in GAT is 12.
In the interest of fairness, we also set the maximum selected pattern number as 12.
Hyper-parameter tuning is time costly on our industrial data so that we use the best combinations obtained from one day transactions.
We implement our methods and run experiments with Python and PyTorch over 8 x A100 NVIDIA GPUs.

\section{Experimental Results}
\label{appendix:result}

Due to the space limit, we only report some results in the main content.
More comprehensive comparisons are shown in Tables~\ref{tab:next_item_diginetica}-\ref{tab:next_item_elec}, where standard deviations are enclosed in brackets.
The best and second-best results are respectively highlighted in bold and underlined.
Methods that use attributes are marked with \ddag, and * indicates the $p$-value $<$ 0.0001 in t-test.

\begin{table*}[t]
    \sisetup{detect-weight=true,detect-inline-weight=math}
    \centering
    \small
    \setlength\tabcolsep{4pt}
    \captionof{table}{Performance evaluation for next-item prediction on \textit{diginetica}.}
	\label{tab:next_item_diginetica}
	\begin{tabular}{l|ccc|ccc}
        \toprule
        \multicolumn{1}{c|}{\multirow{2}{*}{Model}} & 
        \multicolumn{6}{c}{diginetica} \\
        & \multicolumn{1}{c}{{Hits@10}} & \multicolumn{1}{c}{{NDCG@10}} & \multicolumn{1}{c|}{{MRR@10}}
        & \multicolumn{1}{c}{{Hits@20}} & \multicolumn{1}{c}{{NDCG@20}} & \multicolumn{1}{c}{{MRR@20}}
        \\
        \midrule
		FPMC & 31.57\scriptsize{(0.04)}* & 17.40\scriptsize{(0.01)}* & 13.08\scriptsize{(0.02)}* & 43.19\scriptsize{(0.05)}* & 20.33\scriptsize{(0.03)}* & 13.88\scriptsize{(0.03)}* \\
		GRU4Rec & \underline{36.77}\scriptsize{(0.14)}* & 20.71\scriptsize{(0.05)}* & 15.80\scriptsize{(0.03)}* & \underline{49.68}\scriptsize{(0.06)}* & \underline{23.97}\scriptsize{(0.03)}* & 16.70\scriptsize{(0.03)}* \\
		NARM & 35.98\scriptsize{(0.10)}* & 20.18\scriptsize{(0.06)}* & 15.36\scriptsize{(0.06)}* & 48.89\scriptsize{(0.12)}* & 23.44\scriptsize{(0.06)}* & 16.26\scriptsize{(0.06)}* \\
		STAMP & 33.59\scriptsize{(0.15)}* & 18.89\scriptsize{(0.18)}* & 14.41\scriptsize{(0.19)}* & 45.87\scriptsize{(0.15)}* & 22.00\scriptsize{(0.18)}* & 15.26\scriptsize{(0.19)}* \\
		CSRM & 33.97\scriptsize{(0.08)}* & 19.43\scriptsize{(0.03)}* & 14.98\scriptsize{(0.03)}* & 45.83\scriptsize{(0.02)}* & 22.42\scriptsize{(0.02)}* & 15.80\scriptsize{(0.02)}* \\
		S3-Rec\ddag & 33.48\scriptsize{(0.13)}* & 18.58\scriptsize{(0.09)}* & 14.04\scriptsize{(0.10)}* & 45.97\scriptsize{(0.08)}* & 21.74\scriptsize{(0.09)}* & 14.90\scriptsize{(0.10)}* \\
		M2TRec\ddag & 29.67\scriptsize{(0.43)}* & 16.30\scriptsize{(0.24)}* & 12.23\scriptsize{(0.18)}* & 41.23\scriptsize{(0.63)}* & 19.22\scriptsize{(0.29)}* & 13.02\scriptsize{(0.20)}* \\
		\hline
		SR-GNN & 35.21\scriptsize{(0.02)}* & 19.68\scriptsize{(0.04)}* & 14.94\scriptsize{(0.04)}* & 47.99\scriptsize{(0.04)}* & 22.90\scriptsize{(0.04)}* & 15.82\scriptsize{(0.04)}* \\
		GC-SAN & 35.25\scriptsize{(0.09)}* & 19.72\scriptsize{(0.04)}* & 14.97\scriptsize{(0.03)}* & 47.87\scriptsize{(0.09)}* & 22.90\scriptsize{(0.04)}* & 15.85\scriptsize{(0.03)}* \\
		S2-DHCN & 30.76\scriptsize{(0.07)}* & 17.04\scriptsize{(0.14)}* & 12.86\scriptsize{(0.16)}* & 42.39\scriptsize{(0.07)}* & 19.98\scriptsize{(0.13)}* & 13.66\scriptsize{(0.16)}* \\
		GCE-GNN & 36.32\scriptsize{(0.09)}* & \underline{20.77}\scriptsize{(0.07)}* & \underline{16.02}\scriptsize{(0.07)}* & 48.67\scriptsize{(1.12)}* & 23.89\scriptsize{(0.23)}* & \underline{16.87}\scriptsize{(0.03)}* \\
		LESSR & 33.68\scriptsize{(0.05)}* & 18.71\scriptsize{(0.03)}* & 14.14\scriptsize{(0.03)}* & 46.23\scriptsize{(0.11)}* & 21.88\scriptsize{(0.05)}* & 15.01\scriptsize{(0.03)}* \\
		MSGIFSR & 34.74\scriptsize{(0.09)}* & 19.43\scriptsize{(0.06)}* & 14.76\scriptsize{(0.07)}* & 46.23\scriptsize{(0.11)}* & 21.88\scriptsize{(0.05)}* & 15.01\scriptsize{(0.03)}* \\
		\hline
		FAPAT\ddag & \textbf{37.42}\scriptsize{(0.10)} & \textbf{21.31}\scriptsize{(0.03)} & \textbf{16.39}\scriptsize{(0.04)} & \textbf{50.41}\scriptsize{(0.15)} & \textbf{24.59}\scriptsize{(0.06)} & \textbf{17.29}\scriptsize{(0.04)} \\
		\textit{Improv.} & \textit{3.03\%} & \textit{2.60\%} & \textit{2.31\%} & \textit{1.46\%} & \textit{2.59\%} & \textit{2.49\%} \\
        \bottomrule
    \end{tabular}
\end{table*}

\begin{table*}[t]
    \sisetup{detect-weight=true,detect-inline-weight=math}
    \centering
    \small
    \setlength\tabcolsep{4pt}
    \captionof{table}{Performance evaluation for next-item prediction on \textit{Tmall}.}
	\label{tab:next_item_tmall}
	\begin{tabular}{l|ccc|ccc}
        \toprule
        \multicolumn{1}{c|}{\multirow{2}{*}{Model}} & 
        \multicolumn{6}{c}{Tmall} \\
        & \multicolumn{1}{c}{{Hits@10}} & \multicolumn{1}{c}{{NDCG@10}} & \multicolumn{1}{c|}{{MRR@10}}
        & \multicolumn{1}{c}{{Hits@20}} & \multicolumn{1}{c}{{NDCG@20}} & \multicolumn{1}{c}{{MRR@20}}
        \\
        \midrule
		FPMC & 13.71\scriptsize{(0.16)}* & 9.02\scriptsize{(0.02)}* & 7.56\scriptsize{(0.03)}* & 16.44\scriptsize{(0.23)}* & 9.71\scriptsize{(0.04)}* & 7.74\scriptsize{(0.02)} \\
		GRU4Rec & 18.82\scriptsize{(0.17)}* & 12.28\scriptsize{(0.11)}* & 10.25\scriptsize{(0.09)}* & 22.68\scriptsize{(0.21)}* & 13.25\scriptsize{(0.12)}* & 10.51\scriptsize{(0.10)}* \\
		NARM & 22.74\scriptsize{(0.20)}* & 15.46\scriptsize{(0.12)}* & 13.19\scriptsize{(0.10)}* & 26.73\scriptsize{(0.26)}* & 16.47\scriptsize{(0.13)}* & 13.47\scriptsize{(0.10)}* \\
		STAMP & 24.32\scriptsize{(0.31)}* & 16.55\scriptsize{(0.29)}* & 14.12\scriptsize{(0.29)}* & 28.40\scriptsize{(0.35)}* & 17.58\scriptsize{(0.30)}* & 14.41\scriptsize{(0.29)}* \\
		CSRM & 25.13\scriptsize{(0.19)}* & 18.56\scriptsize{(0.18)}* & 16.48\scriptsize{(0.18)}* & 27.94\scriptsize{(0.15)}* & 19.27\scriptsize{(0.17)}* & 16.68\scriptsize{(0.18)}* \\
		S3-Rec\ddag & 18.24\scriptsize{(0.11)}* & 12.30\scriptsize{(0.07)}* & 10.46\scriptsize{(0.06)}* & 22.31\scriptsize{(0.17)}* & 13.32\scriptsize{(0.08)}* & 10.74\scriptsize{(0.06)}* \\
		M2TRec\ddag & 11.42\scriptsize{(0.21)}* & 7.56\scriptsize{(0.06)}* & 6.36\scriptsize{(0.11)}* & 13.75\scriptsize{(0.35)}* & 8.15\scriptsize{(0.04)}* & 6.52\scriptsize{(0.10)}* \\
		\hline
		SR-GNN & 18.21\scriptsize{(0.51)}* & 12.11\scriptsize{(0.32)}* & 10.20\scriptsize{(0.28)}* & 21.34\scriptsize{(0.49)}* & 12.91\scriptsize{(0.31)}* & 10.42\scriptsize{(0.28)}* \\
		GC-SAN & 19.29\scriptsize{(0.14)}* & 12.80\scriptsize{(0.07)}* & 10.78\scriptsize{(0.13)}* & 23.18\scriptsize{(0.23)}* & 13.78\scriptsize{(0.04)}* & 11.05\scriptsize{(0.12)}* \\
		S2-DHCN & 22.00\scriptsize{(0.36)}* & 13.36\scriptsize{(0.21)}* & 10.68\scriptsize{(0.17)}* & 27.23\scriptsize{(0.33)}* & 14.69\scriptsize{(0.20)}* & 11.05\scriptsize{(0.17)}* \\
		GCE-GNN & \underline{28.33}\scriptsize{(0.13)}* & \underline{20.01}\scriptsize{(0.12)}* & \underline{17.32}\scriptsize{(0.13)}* & \underline{30.24}\scriptsize{(0.16)}* & \underline{20.50}\scriptsize{(0.13)}* & \underline{17.45}\scriptsize{(0.13)}* \\
		LESSR & 20.99\scriptsize{(0.26)}* & 14.64\scriptsize{(0.18)}* & 12.13\scriptsize{(0.19)}* & 25.92\scriptsize{(0.23)}* & 13.96\scriptsize{(0.22)}* & 10.50\scriptsize{(0.23)}* \\
		MSGIFSR & 23.18\scriptsize{(0.19)}* & 15.19\scriptsize{(0.11)}* & 12.69\scriptsize{(0.10)}* & 27.78\scriptsize{(0.25)}* & 16.35\scriptsize{(0.11)}* & 13.01\scriptsize{(0.09)}* \\
		\hline
		FAPAT\ddag & \textbf{32.45}\scriptsize{(0.21)} & \textbf{22.02}\scriptsize{(0.15)} & \textbf{18.72}\scriptsize{(0.13)} & \textbf{36.18}\scriptsize{(0.21)} & \textbf{22.97}\scriptsize{(0.14)} & \textbf{18.99}\scriptsize{(0.13)} \\
		\textit{Improv.} & \textit{14.19\%} & \textit{10.04\%} & \textit{8.08\%} & \textit{19.64\%} & \textit{12.05\%} & \textit{8.83\%} \\
        \bottomrule
    \end{tabular}
\end{table*}

\begin{table*}[t]
    \sisetup{detect-weight=true,detect-inline-weight=math}
    \centering
    \small
    \setlength\tabcolsep{4pt}
    \captionof{table}{Performance evaluation for next-item prediction on \textit{Beauty}.}
	\label{tab:next_item_beauty}
	\begin{tabular}{l|ccc|ccc}
        \toprule
        \multicolumn{1}{c|}{\multirow{2}{*}{Model}} & 
        \multicolumn{6}{c}{Beauty} \\
        & \multicolumn{1}{c}{{Hits@10}} & \multicolumn{1}{c}{{NDCG@10}} & \multicolumn{1}{c|}{{MRR@10}}
        & \multicolumn{1}{c}{{Hits@20}} & \multicolumn{1}{c}{{NDCG@20}} & \multicolumn{1}{c}{{MRR@20}}
        \\
        \midrule
		FPMC & 72.00 & 57.20 & 52.42 & 75.91 & 58.19 & 52.70 \\
		GRU4Rec & 73.95 & 58.19 & 53.13 & 78.54 & 59.36 & 53.45 \\
		NARM & 88.09 & 70.44 & 64.68 & 91.50 & 71.31 & 64.93 \\
		STAMP & 80.08 & 63.76 & 58.47 & 83.84 & 64.72 & 58.73 \\
		CSRM & 89.74 & 75.28 & 70.56 & 92.61 & 76.01 & 70.77 \\
		S3-Rec\ddag & 89.64 & \underline{75.56} & \underline{70.99} & 92.53 & \underline{76.30} & \underline{71.19} \\
		M2TRec\ddag & 80.13 & 65.97 & 61.65 & 83.66 & 66.87 & 61.65 \\
		\hline
		SR-GNN & 88.69 & 70.42 & 64.44 & 91.74 & 71.20 & 64.65 \\
		GC-SAN & 86.67 & 70.80 & 64.71 & 88.98 & 72.50 & 65.97 \\
		S2-DHCN & 7.25 & 5.38 & 4.80 & 8.87 & 5.79 & 4.91 \\
		GCE-GNN & 89.34 & 73.15 & 67.80 & 91.29 & 73.65 & 67.94 \\
		LESSR & 89.95 & 71.29 & 65.18 & \underline{92.98} & 72.06 & 65.40 \\
		MSGIFSR & \underline{90.18} & 73.62 & 65.18 & 92.50 & 74.21 & 65.65 \\
		\hline
		FAPAT\ddag & \textbf{92.72} & \textbf{76.29} & \textbf{71.09} & \textbf{94.10} & \textbf{76.87} & \textbf{71.24} \\
		\textit{Improv.} & \textit{2.82\%} & \textit{0.97\%} & \textit{0.14\%} & \textit{1.20\%} & \textit{0.75\%} & \textit{0.07\%} \\
        \bottomrule
    \end{tabular}
\end{table*}

\begin{table*}[t]
    \sisetup{detect-weight=true,detect-inline-weight=math}
    \centering
    \small
    \setlength\tabcolsep{4pt}
    \captionof{table}{Performance evaluation for next-item prediction on \textit{Books}.}
	\label{tab:next_item_book}
	\begin{tabular}{l|ccc|ccc}
        \toprule
        \multicolumn{1}{c|}{\multirow{2}{*}{Model}} & 
        \multicolumn{6}{c}{Books} \\
        & \multicolumn{1}{c}{{Hits@10}} & \multicolumn{1}{c}{{NDCG@10}} & \multicolumn{1}{c|}{{MRR@10}}
        & \multicolumn{1}{c}{{Hits@20}} & \multicolumn{1}{c}{{NDCG@20}} & \multicolumn{1}{c}{{MRR@20}}
        \\
        \midrule
		FPMC & 36.51 & 24.32 & 20.49 & 41.90 & 25.69 & 20.87 \\
		GRU4Rec & 47.21 & 31.86 & 27.02 & 53.55 & 33.47 & 27.46 \\
		NARM & 76.09 & 54.22 & 47.13 & 80.83 & 55.43 & 47.36 \\
		STAMP & 61.49 & 42.13 & 35.95 & 67.46 & 43.65 & 36.37 \\
		CSRM & \underline{78.69} & 56.70 & 49.54 & \underline{82.88} & 57.77 & 49.83 \\
		S3-Rec\ddag & 75.00 & \underline{58.54} & \underline{53.23} & 79.45 & \underline{59.67} & \underline{53.55} \\
		M2TRec\ddag & 32.56 & 22.58 & 24.98 & 35.39 & 25.70 & 22.78 \\
		\hline
		SR-GNN & 66.55 & 47.55 & 41.32 & 69.77 & 48.37 & 41.55 \\
		GC-SAN & 72.56 & 54.92 & 49.25 & 75.73 & 56.05 & 50.14 \\
		S2-DHCN & 4.69 & 3.42 & 3.03 & 5.60 & 3.65 & 3.09 \\
		GCE-GNN & 77.61 & 57.60 & 51.00 & 80.03 & 58.22 & 51.17 \\
		LESSR & 73.72 & 53.86 & 47.36 & 82.31 & 54.77 & 47.61 \\
		MSGIFSR & 72.93 & 52.23 & 45.66 & 76.33 & 53.09 & 45.66 \\
		\hline
		FAPAT\ddag & \textbf{81.62} & \textbf{61.08} & \textbf{54.39} & \textbf{85.12} & \textbf{61.97} & \textbf{54.64} \\
		\textit{Improv.} & \textit{3.72\%} & \textit{4.34\%} & \textit{2.18\%} & \textit{2.70\%} & \textit{3.85\%} & \textit{2.04\%} \\
        \bottomrule
    \end{tabular}
\end{table*}

\begin{table*}[t]
    \sisetup{detect-weight=true,detect-inline-weight=math}
    \centering
    \small
    \setlength\tabcolsep{4pt}
    \captionof{table}{Performance evaluation for next-item prediction on \textit{Electronics}.}
	\label{tab:next_item_elec}
	\begin{tabular}{l|ccc|ccc}
        \toprule
        \multicolumn{1}{c|}{\multirow{2}{*}{Model}} & 
        \multicolumn{6}{c}{Electronics} \\
        & \multicolumn{1}{c}{{Hits@10}} & \multicolumn{1}{c}{{NDCG@10}} & \multicolumn{1}{c|}{{MRR@10}}
        & \multicolumn{1}{c}{{Hits@20}} & \multicolumn{1}{c}{{NDCG@20}} & \multicolumn{1}{c}{{MRR@20}}
        \\
        \midrule
		FPMC & 37.87 & 26.91 & 23.42 & 42.07 & 27.97 & 23.71 \\
		GRU4Rec & 58.46 & 40.69 & 35.02 & 64.42 & 42.21 & 35.44 \\
		NARM & 61.10 & 41.20 & 32.05 & 77.36 & 44.56 & 33.75 \\
		STAMP & 59.30 & 42.04 & 36.53 & 67.94 & 45.07 & 36.97 \\
		CSRM & 62.28 & 44.35 & 38.59 & 67.47 & 45.67 & 38.96 \\
		S3-Rec\ddag & 74.36 & \underline{56.03} & \textbf{50.16} & 79.63 & \underline{57.37} & \textbf{50.53} \\
		M2TRec\ddag & 57.32 & 44.84 & 40.85 & 61.70 & 45.95 & 41.15 \\
		\hline
		SR-GNN & \underline{74.86} & 54.30 & 47.66 & \underline{79.66} & 55.52 & 48.00 \\
		GC-SAN & 72.76 & 53.37 & 45.98 & 77.34 & 46.34 & 49.91 \\
		S2-DHCN & 4.18 & 2.65 & 2.18 & 5.08 & 2.88 & 2.24 \\
		GCE-GNN & 72.93 & 53.74 & 47.59 & 78.49 & 55.15 & 47.98 \\
		LESSR & 72.91 & 50.46 & 43.26 & 78.78 & 51.96 & 43.67 \\
		MSGIFSR & 73.56 & 53.83 & 47.77 & 77.45 & 54.73 & 48.02 \\
		\hline
		FAPAT\ddag & \textbf{78.36} & \textbf{56.81} & \underline{49.80} & \textbf{82.81} & \textbf{57.94} & \underline{50.12} \\
		\textit{Improv.} & \textit{4.68\%} & \textit{1.39\%} & \textit{-0.07\%} & \textit{3.95\%} & \textit{0.99\%} & \textit{-0.81\%} \\
        \bottomrule
    \end{tabular}
\end{table*}

\end{document}